\documentclass[pra,aps,preprint,showpacs]{revtex4-2}

\usepackage{amssymb}
\usepackage{amsmath}
\usepackage{amsfonts}
\usepackage{graphicx}
\usepackage{color}
\usepackage[table,xcdraw]{xcolor}
\usepackage{appendix}
\usepackage{bm}
\usepackage[sort&compress]{natbib}
\usepackage{float}

\begin{document}

\title{Time-dependent theory of reconstruction of attosecond harmonic beating by interference of multi-photon transitions}
\author{Sebasti\'an D. L\'{o}pez}
\email{sebastian.lopez@conicet.gov.ar}
\affiliation{Instituto de Investigaciones en Energ\'ia No Convencional - INENCO (UNSa - CONICET), Av. Bolivia 5150, 4400, Salta Capital, Argentina}
\author{Matias L. Ocello}
\affiliation{Institute for Astronomy and Space Physics - IAFE (UBA-CONICET), C1428GA, Buenos Aires, Argentina}
\author{Diego G. Arb\'{o}}
\affiliation{Institute for Astronomy and Space Physics - IAFE (UBA-CONICET), C1428GA, Buenos Aires, Argentina}
\affiliation{Universidad de Buenos Aires, Facultad de Ciencias Exactas y Naturales y
Ciclo B\'{a}sico Com\'{u}n, Buenos Aires, Argentina.}

\date{\today }

\begin{abstract}

Phase and time delays of atomic above-threshold ionization are usually experimentally explored by the reconstruction of attosecond harmonic beating by interference of two-photon transitions (RABBIT) technique. Theoretical studies of RABBIT rely on the perturbative treatment of the probe (NIR or visible) laser pulse with respect to the atomic electric field and the  pump composed of a train of attosecond pulses made of several harmonics with frequencies multiple of the probe fundamental frequency. 
In this work we present a semiclassical non-perturbative description of the phase delays for the emission of electrons from hydrogen atoms based on the strong-field approximation as the relative phase between pump and probe pulses is varied, where more than two photons are involved.
Ionization times are calculated within the saddle-point approximations and serve to individualize the different electron wave packets that produce the RABBIT interferometric scheme. We observe different behaviors of the phase delays at different intensities  of the probe.
For example, for moderate and intense probe fields, the harmonics and sidebands happen to be in phase ($ \gtrsim 4 \times 10^{11}$  W/cm$^2$). In turn, when the probe field is sufficiently weak, we recover the well-known rule of thumb for the phase delays developed within the perturbative RABBIT theory [see D. Gu\'enot \textit{et al}. Phys. Rev. A \textbf{85}, 053424 (2012)]. We show that the intracycle interference of the different paths contributing to the final energy (sideband or high harmonic) is responsible for the different behaviors of the interference pattern. 
Comparisons with the numerical 
solution of the strong-field approximation and time-dependent Schr\"odinger equation
confirm the reliability of our semiclassical non-perturbative theory. 

\end{abstract}


\maketitle

\section{Introduction \label{introduction}}

One of the major advances of the applications in attosecond pulses and phase-controlled femtosecond laser pulses consists of measuring and analyzing atomic, molecular \cite{Huppert16,Ning14}, and surface \cite{Cavalieri07,Neppl12,Ossiander18} ionization by absorption of an XUV photon to retrieve timing information of the process encoded in wavepacket phases \cite{Krausz09,Pazourek15}. 
The shortest timescales have been accessed through the combination of XUV pulses and near-infrared or visible lasers \cite{Veniard95,Schins96,Glover96,Hummert20},
which serves as the cornerstone of attosecond chronoscopy of photoionization processes. 
To measure attosecond time-resolved electron emission from noble gas atoms \cite{Schultze10,Klunder11,Guenot12, Guenot14, Dahlstrom13, Fuchs20}, molecules \cite{Huppert16,Beaulieu17}, and solids \cite{Cavalieri07,Lemell15,Haessler15}
several pump-probe techniques have been used, such as attosecond streaking \cite{Itatani02, Goulielmakis04,Goulielmakis08} and reconstruction of attosecond harmonic beating by interference of two-photon transitions (RABBIT) \cite{Veniard95,Veniard96,Paul01}. Whereas the former can be understood classically as the electrons ionized by an XUV pulse are subsequently shifted in energy by the probe laser \cite{Itatani02,Nagele12,Pazourek12,DellaPicca20a,DellaPicca20b,Dahlstrom12}, the latter is comprehended as a quantum interferometer resulting from two interfering paths to the same final state in the continuum \cite{Klunder11,Dahlstrom12}.
This final state can be reached through a two-photon process involving the absorption or emission of an IR photon of the fundamental driving frequency following the absorption of photons from a train of pulses formed by two adjacent (odd) harmonic orders of high-order harmonic generation (HHG) radiation \cite{Kheifets13,Feist14,Su13}.  
Most of the theoretical studies of the RABBIT technique have been performed for low-intensity probe fields in the perturbative limit, such that it does not modify the ionization process appreciably. However, continuum-continuum transitions produced by the probe pulse are inherent to the measurement process in RABBIT and are unavoidable to any experimental setup. The problem of the measurement of the continuum-continuum transitions can be solved by using exact calculations in the perturbative regime.

In this paper, we theoretically investigate the phase delays for the photoelectron emission from atomic hydrogen in a RABBIT-like protocol beyond the perturbative limit of the probe field. A similar non-perturbative study was recently developed for atomic above-threshold ionization by a pulse composed of the fundamental frequency and its first harmonic \cite{Lopez22}. We develop a semiclassical model considering different ionization times calculated within the saddle-point approximation, which is a generalization of the theory for diffraction at a time grating \cite{Arbo10a,Arbo10b,Arbo12}. Our model, based on the strong-field approximation (SFA) considers all possible transition paths to the final energy state containing multiple/infinite probe photons and not just two-photon transitions as in the perturbative limit.
Within our semiclassical model (SCM) and, generally, the SFA, the influence of the Coulomb potential of the remaining core on the outgoing electron wavepacket is neglected \cite{Faria99,Arbo10b,Lai15,Faria20,Maxwell20}. We find and report substantially different behaviors of the sidebands and harmonics as a function of the relative phase between the pump and probe pulses from the well-established perturbative regime for probe intensities of the order of $10^{11}-10^{12}$ W/cm$^2$. We compare the SCM with the results of the numerical solution of the SFA and the time-dependent Schr\"odinger equation (TDSE), indicating the reliability of our model and the minor effect of the Coulomb potential of the remaining core on the emitted electron wavepacket for the atomic and laser parameters used \cite{Faria99,Faria20,Arbo10b,Nagele12,Pazourek12,Kheifets13,Su13,Feist14,Pazourek15,Lai15,Boll16,Maxwell20,boll22}. 
We conclude that the intracycle interference of electron trajectories mimicking the emitted wave packets \cite{Arbo10a,Arbo10b,Arbo12} plays a crucial role to determine the phase delays for atomic photoionization in the RABBIT protocol.

The structure of the paper is as follows. In Sec. \ref{theory} we develop the semiclassical theory previously used in laser-assisted photoionization emission (LAPE) considering only one color pump field \cite{Gramajo16,Gramajo17,Gramajo18} and then extend it to the case of a two-color pump pulse train. There, we show how to extract the information on ionization phases and phase delays. In Sec.~\ref{results} we present numerical results for energy spectra calculated with our SCM and compare them with the results of the SFA and the TDSE. We conclude our remarks in Sec. \ref{conclusions}. Atomic units ($e=\hbar =m_{e}=1$ a.u.) are used throughout unless stated otherwise.


\section{Semiclassical theory \label{theory}}

We consider the ionization of an atomic system by linearly polarized laser
pulses. In the single-active-electron (SAE) approximation, the TDSE reads 
\begin{equation}
i\frac{\partial }{\partial t}\left\vert \psi (t)\right\rangle =\left[
H_{0}+H_{\text{int}}(t)\right] \left\vert \psi (t)\right\rangle ,
\label{TDSE}
\end{equation}%
where $H_{0}=\vec{p}^{2}/2+V(r)$ is the time-independent atomic Hamiltonian,
whose first term corresponds to the electron kinetic energy and its second
term to the electron-core Coulomb interaction and $H_{\text{int}}(t)$
is the interaction Hamiltonian between the atomic system
and the external laser field. Because of the presence of the external laser field, the initially bound electron in an atomic state $\vert \phi _{i}\rangle $ is emitted with final
momentum $\vec{k}$ and energy $E=k^{2}/2$ in a final continuum state 
$\vert \phi_{f}\rangle$.
Because of the azimuthal symmetry, the electron probability distribution can be expressed in terms of only two variables, i.e.,
the electron kinetic energy $E$ and the polar electron emission angle $%
\theta $, or equivalently, the electron momentum parallel $k_{z}$ and
transversal $k_{\rho }$ to the field polarization direction, i.e., 
\begin{equation*}
\left\vert T\right\vert ^{2}=\frac{dP}{2\pi \sqrt{2E}\ dE\ d\cos \theta }=%
\frac{dP}{2\pi k_{\rho }dk_{\rho }dk_{z}}.
\label{proba}
\end{equation*}

Within the time-dependent distorted wave theory, the
transition amplitude in the prior form and length gauge is expressed as \cite{Macri03,Arbo08a}
\begin{equation}
T=-i\int_{-\infty }^{+\infty }dt\,\langle \chi _{f}^{-}(\vec{r},t)|H_{\text{%
int}}(\vec{r},t)|\phi _{i}(\vec{r},t)\rangle ,  
\label{Tif}
\end{equation}
where $\phi _{i}(\vec{r},t)=\varphi _{i}(\vec{r})\,e^{iI_{p}t}$ is the
initial atomic state with ionization potential $I_{p}$ and $\chi _{f}^{-}(%
\vec{r},t)$ is the distorted final state. Eq. (\ref{Tif}) is exact provided
the final channel $\chi _{f}^{-}(\vec{r},t)$ is the exact solution of Eq. (%
\ref{TDSE}). Several degrees of approximation have been considered in the
literature to solve Eq. (\ref{Tif}). The widest known one is the SFA,
which neglects in the final channel the Coulomb
distortion produced on the ejected-electron state due to its interaction
with the residual ion and discards the influence of the laser field in the
initial ground state~\cite{Lewenstein94,Lewenstein95}. Hence, the SFA consists
of approximating the distorted final state with the solution of the TDSE for
a free electron in an electromagnetic field, namely, a Volkov function \cite%
{Volkov35}, i.e., $\chi _{f}^{-}(\vec{r},t)=\chi _{f}^{V}(\vec{r},t)$, where 
\begin{eqnarray}
\chi _{f}^{V}(\vec{r},t) &=&\frac{1}{(2\pi )^{3/2}}\exp \{i[\vec{k}+\vec{A}%
(t)]\cdot \vec{r}\}  \notag \\
&\times &\exp \left\{ \frac{i}{2}\int_{t}^{\infty }[\vec{k}+\vec{A}%
(t^{\prime })]^{2}dt^{\prime }\right\}   \label{Volkov}
\end{eqnarray}%
and the vector potential due to the total external field is defined as $\vec{%
A}(t)=-\int_{-\infty }^{t}dt^{\prime }\vec{F}(t^{\prime })$. 

We consider the following expression for a laser pulse given by an electric field
with main frequency $\omega $, and its odd harmonics $\left( 2m+1\right) \omega $, with $m=1,2,3,...$
\begin{equation}
\vec{F}(t)=f(t)\left[ F_{0}\cos (\omega t+\phi
)+ \sum_{m=1}^{\infty }F_{m}\cos \left[ \left( 2m+1\right) \omega
t+\phi _{m}\right] \right] \hat{z},  \label{field1}
\end{equation}
where $\phi $ is the relative phase of the fundamental laser field with respect
to its harmonics, $f(t)$ is a smooth function between $0$ and $1$
mimicking the pulse envelope, $\hat{z}$ is the polarization direction of
both fundamental and odd harmonic fields, and $F_{m}$ are the field strengths
with the special case of $m=0$ for the fundamental frequency.
Each harmonic undertakes a phase $\phi_m$.
For a long pulse with
adiabatic switch on and off the vector potential in the length gauge
can be written in its central part as 
\begin{subequations}
\label{Avector}
\begin{eqnarray}
\vec{A}(t) &\simeq &-\frac{f(t)}{\omega } \Bigl\{ F_{0}\sin (\omega t + \phi ) + \sum_{m=1}^{\infty }\frac{F_{m}}{%
\left( 2m+1\right) }\sin [ (2m+1)\omega t+\phi _{m} ]  \Bigl\} \hat{z}  \\
&\simeq &-\frac{f(t)}{\omega }F_{0}\sin (\omega t + \phi)\ \hat{z},  
\end{eqnarray}
\end{subequations}
where in Eq. (\ref{Avector}b) we have neglected the
contribution of all the harmonics ($m=1,2,...$) since we consider either 
high harmonic orders, i.e., $m\gg 1$ and/or low harmonic intensity compared to the
intensity of the fundamental laser field, i.e., $F_{m}\ll F_{0}(2m+1)$.
In the high-energy region of the spectrum, we can consider this approximation
to be very accurate, however, for ionization near threshold it is dubious.

Let us now analyze some features of the ionization amplitude in Eq. (\ref{Tif}).
Using Eq. (\ref{field1}), the transition matrix $T$ in Eq. (\ref{Tif}) is
separable in the different harmonics of the laser fields, i.e., 
$T=\sum_{m}T_{m}$, with
\begin{equation}
T_{m}=F_{m}\int_{-\infty }^{+\infty }\,l(t)\ e^{i\left[ S_m(t)-\phi _{m}%
\right] }\,\,dt,  
\label{Tm}
\end{equation}
where
\begin{equation}
\ell (t)=-\frac{i}{2}f(t)\ \hat{z}\cdot \vec{d}\left[ \vec{k}+\vec{A}(t)%
\right]  ,
\label{dipole-factor}
\end{equation}
and
\begin{equation}
S_m(t)=-\int_{t}^{\infty }dt^{\prime }\left\{ \frac{\left[ \vec{k}+\vec{A}%
(t^{\prime })\right] ^{2}}{2}+I_{p}-(2m+1)\omega \right\} ,  
\label{action}
\end{equation}
with the dipole moment defined as $\vec{d}(\vec{v})=(2\pi )^{-3/2}\langle
e^{i\vec{v}\cdot \vec{r}}|\vec{r}|\varphi _{i}(\vec{r})\rangle $, and $S_m(t)$
is the generalized classical action for the harmonic frequency $(2m+1)\omega$.
In Eq. (\ref{Tm}) we have considered the high
frequency electric field within the rotating-wave approximation \cite%
{Gramajo16,Gramajo17,Gramajo18}. This approximation considers the
high-frequency time-dependent ionization amplitude due to the high-harmonic
as a single photon absorption transition of energy $(2m+1)\omega $.
We restrict our analysis to the energy domain where ionization due to the laser of fundamental frequency is negligible.

Considering the vector potential in Eq. (\ref{Avector}b) in the
flat-top region, i.e., $f(t)=1,$ the time integration of the action $S(t)$
defined in Eq. (\ref{action}) can be analytically written as 
\begin{equation}
S_m(t)=at+b\cos (\omega t + \phi)+c\sin (2\omega t + 2 \phi), \label{action2}
\end{equation}%
where
\begin{subequations}
\label{abc}
\begin{eqnarray}
a &=&\frac{k^{2}}{2}+I_{p}+U_{p}-(2m+1)\omega , \\
b &=&\alpha k_{z}, \\
c &=&-\frac{U_{p}}{2\omega },
\end{eqnarray}
\end{subequations}
$\alpha =F_{0}/\omega ^{2}$ defines the quiver amplitude of the electron and 
$U_{p}=(F_{0}/2\omega )^{2}$ defines its ponderomotive energy under the
electric field of fundamental frequency \footnote{In Eq. (\ref{action2}) we have omitted the constant term $U_P / \omega \phi$ since it has null contribution to the transition matrix.}.

The vector potential in Eq. (\ref{Avector}b) together with the dipole moment result to
be $T$-periodic, i.e., $\vec{d}\left[ \vec{k}+\vec{A}(t+jT)\right] =\vec{d}%
\left[ \vec{k}+\vec{A}(t)\right] ,$ with $j$ any positive integer number if
we consider that the central (and not varying) part of the field starts at $%
t=0.$ The reader must not confuse the fundamental laser period $T=2\pi
/\omega $ with the transition amplitude $T$, although we use the same
symbol. From Eq. (\ref{action2}), we
observe that $[S_m(t)-at]$ is a time-oscillating function with the same period 
$T$ of the fundamental laser field, i.e., 
\begin{equation}
S_m(t+jT)=S_m(t)+ajT.  \label{S-periodic}
\end{equation}
In light of these periodicity properties [Eq. (\ref{S-periodic}) and $\ell
(t+jT)=\ell (t)$ in Eq. (\ref{dipole-factor})], 
we can rewrite the transition matrix $T_{m}$ in Eq. (\ref{Tm})
in terms of the contribution of the first fundamental cycle just
assuming that the (fundamental) probe field is composed of $N$ optical
cycles each of duration $T$, and neglecting the contributions
of the ramp on and ramp off, i.e.,
\begin{eqnarray}
T_{m} &=&F_{m}\int_{0}^{NT}\,\ell (t)e^{i\left[ S_m(t)-\phi _{m}\right]
}\,\,dt  \notag \\
&=&F_{m}e^{-i\phi _{m}}\sum_{j=0}^{N-1}\int_{jT}^{(j+1)T}\ell
(t)e^{iS_m(t)}dt.  \label{Tm1}
\end{eqnarray}
By performing the transformation $t=t^{\prime }+jT$, the temporal
integral into the second line of Eq. (\ref{Tm1}) becomes delayed in $j$
cycles of the probe laser. Keeping in mind the $T$-periodicity of both $\ell
(t)$ and $S_m(t)-at$ [see Eq.~(\ref{S-periodic})], it is straightforward to
factorize the transition amplitude as 
\begin{eqnarray}
T_{m} &=&F_{m}e^{-i\phi _{m}}\sum_{j=0}^{N-1}e^{iajT}\int_{0}^{T}\ell
(t^{\prime })e^{iS_m(t^{\prime })}dt^{\prime }  \notag \\
&=&F_{m}e^{-i\phi_{m}}\frac{\sin {(aTN/2)}}{\sin {(aT/2)}}%
\,e^{(iaT(N-1)/2)}I_{m}(\vec{k}),  \label{Tm3}
\end{eqnarray}
where the factor
\begin{equation}
I_{m}(\vec{k})=\int_{0}^{T}\ell (t^{\prime })e^{iS_m(t^{\prime })}dt^{\prime }  \label{Imk}
\end{equation}
in Eq. (\ref{Tm3}) corresponds to the contribution of ionization amplitude into one
optical cycle of the fundamental field and its squared absolute value 
$|I_{m}(\vec{k})|^{2}$ is known in the literature as the intracycle
contribution to the ionization probability in laser-asisted photoionization
emission (LAPE) \cite{Arbo08a,Arbo08b,DellaPicca20}.
Thus, the photoelectron spectrum (PES) and
ionization probability [Eq. (\ref{Tm3})] can be expressed as a product
of the intracycle factor $|I_{m}(\vec{k})|^{2}$ and the factor $|\sin 
{(aTN/2)}/\sin {(aT/2)}|^{2}$ accounting for the \emph{intercycle}
contributions, since it is the result of the phase interference arising from
the $N$ different optical cycles of the probe field \cite{Arbo10a,Arbo10b,Arbo12}. We want to point out that Eq. (\ref{Tm3}) is a mere consequence of the periodicity of
the transition matrix within the SFA with no further approximations 
\cite{DellaPicca13,DellaPicca20}.

The zeros of the denominator in the intercycle factor Eq. (\ref{Tm3}), i.e., 
the energy values satisfying $aT/2=n\pi $, are avoidable singularities since the
numerator also cancels out and finite maxima are reached at these points.
They occur when
\begin{equation}
E_{2m+1+n}=\left( 2m+1+n\right) \omega -I_{p}-U_{p},  
\label{SB}
\end{equation}%
corresponding to the absorption of one photon of frequency $\left(
2m+1\right) \omega $ followed by the absorption or emission of $|n|$ photons
of fundamental frequency $\omega $ when $n$ is positive or negative,
respectively. Such maxima at energies given by Eq. (\ref{SB}) are recognized
as the sidebands in the PES of the $\left( 2m+1\right) $-th harmonic in
presence of the fundamental laser \cite{Gramajo16,Gramajo17,Gramajo18}. In
fact, when $N\rightarrow \infty $, the intercycle factor becomes a series of
Dirac delta functions, i.e., $\sum_{n}\delta (E-E_{2m+1+n})$, satisfying the
conservation of energy. Instead, for a finite pump pulse of duration $\tau $ (of
the order of $NT$), each sideband has a width $\Delta E\sim 2\pi /NT = \omega / N$,
fulfilling the uncertainty relation $\Delta E\tau \sim 2\pi$, where $\tau
=NT$ is the duration of the pulse.


\subsection{Laser-assisted photoionization emission by one harmonic}

In this subsection we briefly describe the semiclassical theory of LAPE, where
the pump field is composed of a single monochromatic harmonic of order $2m+1$
accompanied by the probe field of fundamental frequency \cite{Gramajo16,
Gramajo18, DellaPicca20}. In order to calculate the transition amplitude
corresponding to the intracycle factor in Eq. (\ref{Imk}) we make use of the
SCM which consists in solving approximately the time integral of the
transition amplitude by means of the saddle-point approximation \cite%
{Chirila05,Corkum94,Ivanov95,Lewenstein95,Gramajo18}. In this sense, the transition amplitude can be approximated as a coherent superposition of the
amplitudes of all electron trajectories with final momentum $\vec{k}$ over the stationary points $t_{s}$ of the generalized action $S_m(t)$ \cite{Gribakin97}, i.e.,
\begin{equation}
I_{m}(\vec{k})\simeq 
\sum_{t_{s}}\frac{-i\sqrt{2\pi }\hat{z}\cdot \vec{d}(\vec{k}+\vec{A}{(t}_{s}{)})}{2 [ -i\ddot{S}(t_{s})]^{1/2}}\exp 
\left[ iS_m(t_{s})\right],
\label{Tsaddle}
\end{equation}
In the following, we analyze the case $k_{z} \equiv k >0$ (forward emission), 
where the saddle equation $\dot{S}_m \equiv dS_m(t)/dt=0$ reads
\begin{equation}
A(t_{s})=\beta _{+}(k)\equiv v_{0,m}-k,  \label{saddle-s}
\end{equation}
with $v_{0,m}^{2}/2=\left( 2m+1\right) \omega -I_{p}$ corresponding to the
energy of photoelectrons ionized by the high-frequency pump pulse in absence of
the probe field of fundamental frequency. Besides, in Eq. (\ref{Tsaddle}),
$\ddot{S}_m(t_{s})\equiv d^{2}S_m(t_{s})/dt^{2}=-\left[ k + A(t_{s})
\right] F_{0} \cos(\omega t_{s} + \phi)$ is independent of the harmonic order index $m$. The solutions of the saddle equation (\ref{saddle-s}) are shown in Appendix \ref{Saddle Times Appendix}.

The transition matrix of equation (\ref{Tsaddle}) in the classical region with real 
ionization times results in 
\begin{equation}
I_{m}(k)=\sum_{\alpha =1}^{2}\,g_{m}(k,t^{(\alpha )})\exp \left[
iS_m(t^{(\alpha )})+i\frac{\pi }{4}\mathrm{sgn}[\ddot{S}_m(t^{(\alpha )})]\right],  \label{Imofk}
\end{equation}
where the weighting factors are
\begin{equation}
g_{m}(k,t^{(\alpha )})=\frac{-i\sqrt{2\pi } d_z\left[k+A(t^{(\alpha )})\right] }{2 v_{0,m}^{1/2} \left\vert F_{0}\cos(\omega t^{(\alpha )} + \phi)\right\vert ^{1/2} } \, ,
\label{gm}
\end{equation}
and $\mathrm{sgn}[\ddot{S}(t^{(1)})]=\mp 1$ and $\mathrm{sgn}[\ddot{S}%
(t^{(2)})]=\pm 1$ when $\beta _{+}(k)\lessgtr 0,$ since 
$\ddot{S}(t^{(\alpha )})= -v_{0,m} F_{0} \cos (\omega t^{(\alpha )}+\phi)$.

To evaluate the action $S_m$ in Eq. (\ref{action2}) at the ionization times,
we consider the accumulated action $\Delta S_m=S_m(t^{(2)})-S_m(t^{(1)})$
between the two ionization times and the average action 
$\bar{S}_m =[S_m(t^{(1)})+S_m(t^{(2)})]/2$ of the two trajectories released in the same
optical cycle. Then, it can be easily shown that the intracycle factor of the transition amplitude in Eq. (\ref{Imofk}) reads
\begin{equation}
I_{m}(k)=2g_{m}(k,t^{(\alpha )})e^{i\bar{S}_m}\cos {\left( \frac{%
\Delta S_m}{2}-\frac{\pi }{4}\text{\textrm{sgn}}[\beta _{+}(k)]\right)} ,
\label{ImkSP}
\end{equation}
since the ionization rate is identical for the two ionization trajectories
within the same optical cycle, i.e., $g_{m}(k,t^{(1)})=g_{m}(k,t^{(2)})$, according to Eq. (\ref{gm}), and 
$\mathrm{sgn}[\ddot{S}_m(t^{(1)})]=-\mathrm{sgn}[\ddot{S}_m(t^{(2)})]=\mathrm{sgn}[\beta _{+}(k)]$.
From Eqs. (\ref{Tm3}) and (\ref{ImkSP}), the ionization probability can be factorized in intercycle and intracycle factors \cite{Gramajo16}
\begin{eqnarray}
\left\vert T_{m}\right\vert ^{2} &=&F_{m}^{2}\left( \frac{\sin {(aTN/2)}}{%
\sin {(aT/2)}}\right) ^{2}\left\vert I_{m}(k)\right\vert ^{2}  \notag
\\
&=&4F_{m}^{2}\left[ g_{m}(k,t^{(1)})\right] ^{2}\left[ \frac{\sin {%
(aTN/2)}}{\sin {(aT/2)}}\right] ^{2}\cos ^{2}{\left( \frac{\Delta S_m}{2}-%
\frac{\pi }{4}\text{\textrm{sgn}}[\beta _{+}(k)]\right) }.
\label{LAPE}
\end{eqnarray}

\begin{figure}[!htbp]
\centering
\includegraphics[width=0.5 \textwidth]{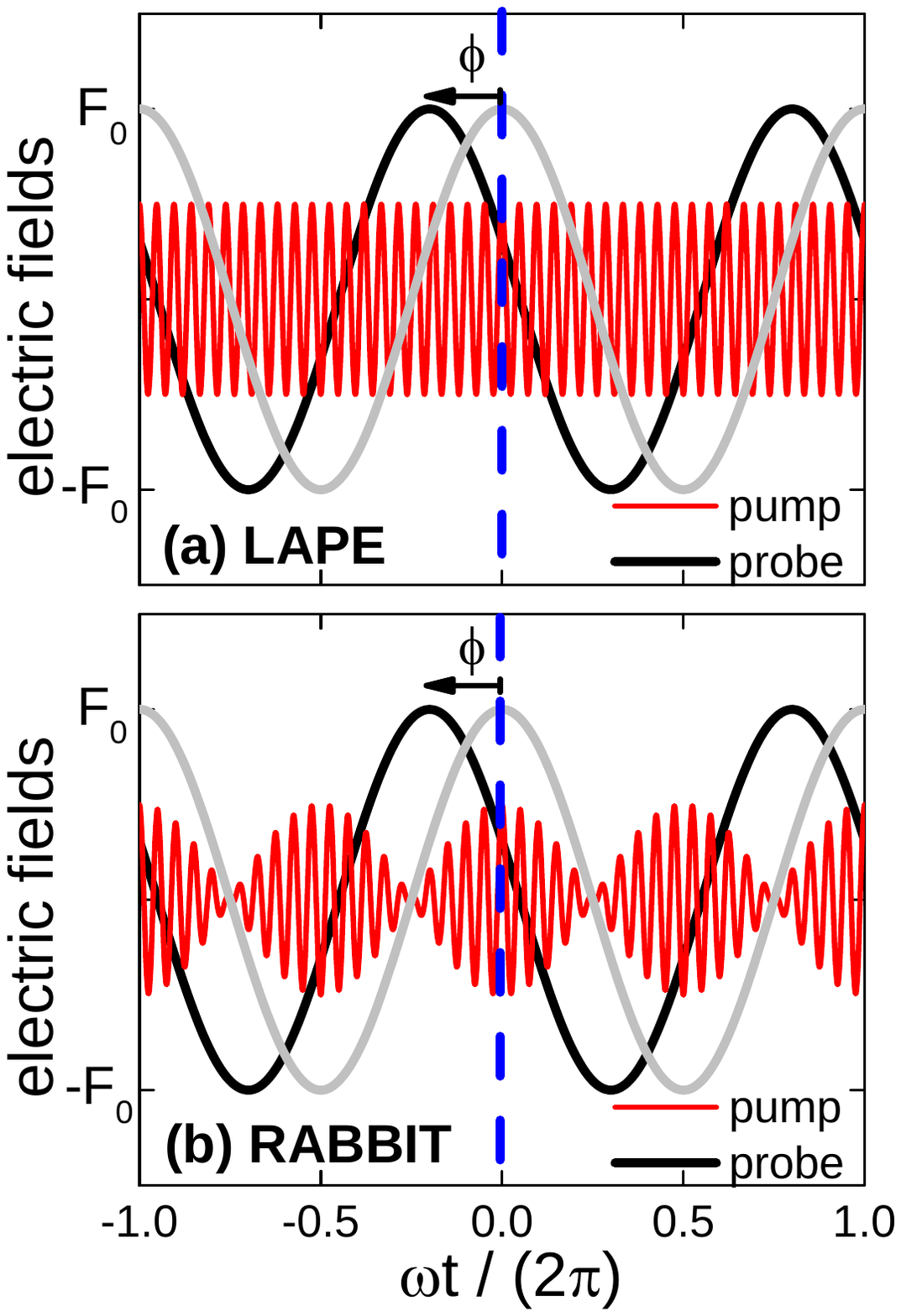}
\caption{Scheme of the electric fields of the probe and the pump as
a function of time. (a) The pump field is monochromatic and thus the
ionization probability is independent of the relative phase $\phi$ of the
probe due to translational invariance in the time domain (for $m \rightarrow \infty$.
(b) When the pump field is composed of various high frequency components (two in this
case), the translational invariance is broken in the time domain (RABBIT).
Thus, the ionization probability depends on the relative phase $\phi$.}
\label{efields}
\end{figure}

Very importantly, in the limit of a pulse of infinite duration, i.e.,  $N \rightarrow \infty$, the ionization probability of the sidebands becomes independent of the relative phase $\phi $ of the probe field [Eq. (\ref{field1})]. The explanation is that (i) the rotating wave approximation used to calculate the transition amplitude in Eq. (\ref{Tm}) considers continuum ionization probability during the XUV pulse duration, which is strictly valid for high harmonics ($m \gg 1$) and (ii) as $N \rightarrow \infty$ any border effect due to the turn on and off of the pulse becomes negligible.
In this sense, there is a translational invariance between the pump and probe pulse as Fig. \ref{efields}a shows. As far as $m$ is very low ($m=1,2$), our assumption is not valid any more, as can be seen in Ref. \cite{Lopez22} for a semiclassical
strong-field theory of phase delays in $\omega - 2\omega$ above threshold ionization.
This invariance can be broken by adding more harmonics as will be shown in the next subsection. 


\subsection{Laser-assisted photoionization emission by two harmonics: Phase delays}

The RABBIT protocol can be thought of as a special case of LAPE, in which a pulse train
formed by several high-harmonic fields is responsible for the ionization of the target.
For the sake of simplicity, we consider atomic ionization by only two neighbor
high-harmonics HH$(m-1)$ and HH$(m)$ of frequencies $(2m-1)\omega $ and $(2m+1)\omega $,
, respectively, followed by continuum-continuum transitions
due to the action of the fundamental frequency $\omega$.
In this sense, the transition probability from the initial atomic bound state to the continuum is given by the coherent superposition of the individual contributions of every HH, i.e.,
\begin{equation}
\left\vert T\right\vert ^{2}=\left\vert T_{m-1}+T_{m}\right\vert ^{2},
\label{T-RABBIT}
\end{equation}
where $T_{m}$ is given by Eq. (\ref{Tm3}).
Unlike LAPE processes with only one HH field contributing to ionization,
in RABBIT the phases of each of the two terms $T_{m-1}$ and $T_{m}$ 
into Eq. (\ref{T-RABBIT}) do matter.
From Fig. \ref{efields}b we observe that in RABBIT, as the pump field is composed of various high-frequency components (two in this case), the translational invariance in the time domain present in ionization by only one HH in Fig. \ref{efields}a is broken. Therefore, one can set an origin for the reference frame of the relative phase of the probe as, for example, the maximum of the envelope of the train of pulses (Fig. \ref{efields}b).
Consequently, the ionization probability [Eq. (\ref{T-RABBIT})]
depends on the relative phase $\phi$, as expected.

Therefore, we focus now on the phase of the probability amplitude for each ATI
(either HH or sideband) peak $n$ [Eq. (\ref{Tm3})].
Considering Eq. (\ref{abc}a) for energies very close to any ATI peak [Eq. (\ref{SB})],
the value of $a=E+I_{p}+U_{p}-(2m+1) \omega \simeq n\omega$ and the total amplitude
in Eq. (\ref{Tm3}) can be written as
\begin{eqnarray}
T_{m}^{(n)} &=&NF_{m}e^{-i\phi _{m}}\frac{\cos {(n\pi N)}}{\cos {(n\pi )}}%
\,e^{in\pi (N-1)}I_{m}(k)  \notag \\
&=&NF_{m}e^{-i\phi _{m}}I_{m}^{(n)},  \label{Tm4}
\end{eqnarray}
where in the first line of Eq. (\ref{Tm4}) we have used the L'H\^{o}pital's
rule for $a\rightarrow n\omega$. If $N$ is even, the factor $\cos {(n\pi N)/%
}\cos {(n\pi )}$ of Eq. (\ref{Tm4}) can be written as $\exp (in\pi )$ and after
multiplicaltion by the factor $e^{in\pi (N-1)}$, the second line of Eq. (\ref{Tm4}) is immediately found since $nN$ is even. On the other hand, if $N$ is odd, the factor $\cos {(n\pi N)/}\cos {(n\pi )}$ of Eq. (\ref{Tm4}) is always equal to the unity and, as $n(N-1)$ is even, the second line of Eq. (\ref{Tm4}) is found again. Therefore, in both cases where the total number of cycles of the fundamental pulse $N$ is even or odd, the phases of the ATI peaks are $ \arg [I_{m}^{(n)}]-\phi_{m}$. In the second line of Eq. (\ref{Tm4}) we have replaced the intracycle amplitude $I_{m}(k)$ at the $n-$th peak by $I_{m}^{(n)}$ making explicit the ATI peak [Eq. (\ref{SB})] order since for each peak the absolute value of the momentum (and energy) is fixed, unlike the emission angle with respect to the polarization axis $\hat{z}$, which we set to zero for forward direction.
For a deeper inspection of the phase of $I_{m-1}^{(n)}$ for each contribution, we calculate the atomic delays within the saddle-point approximation based on the SFA from Eq. (\ref{ImkSP}). For a particular ATI peak, we can rename the quantity $\beta _{+}(k)$ in Eq. (\ref{saddle-s}) as $\beta _{+,m}^{(n)}=\sqrt{2\left[ \left(2m+1\right) \omega -I_{p}\right] }-\sqrt{2\left[ n\omega +\left( 2m+1\right) \omega -I_{p}-U_{p}\right] }$. 



\begin{figure}[!htbp]
\includegraphics[width=0.8\textwidth]{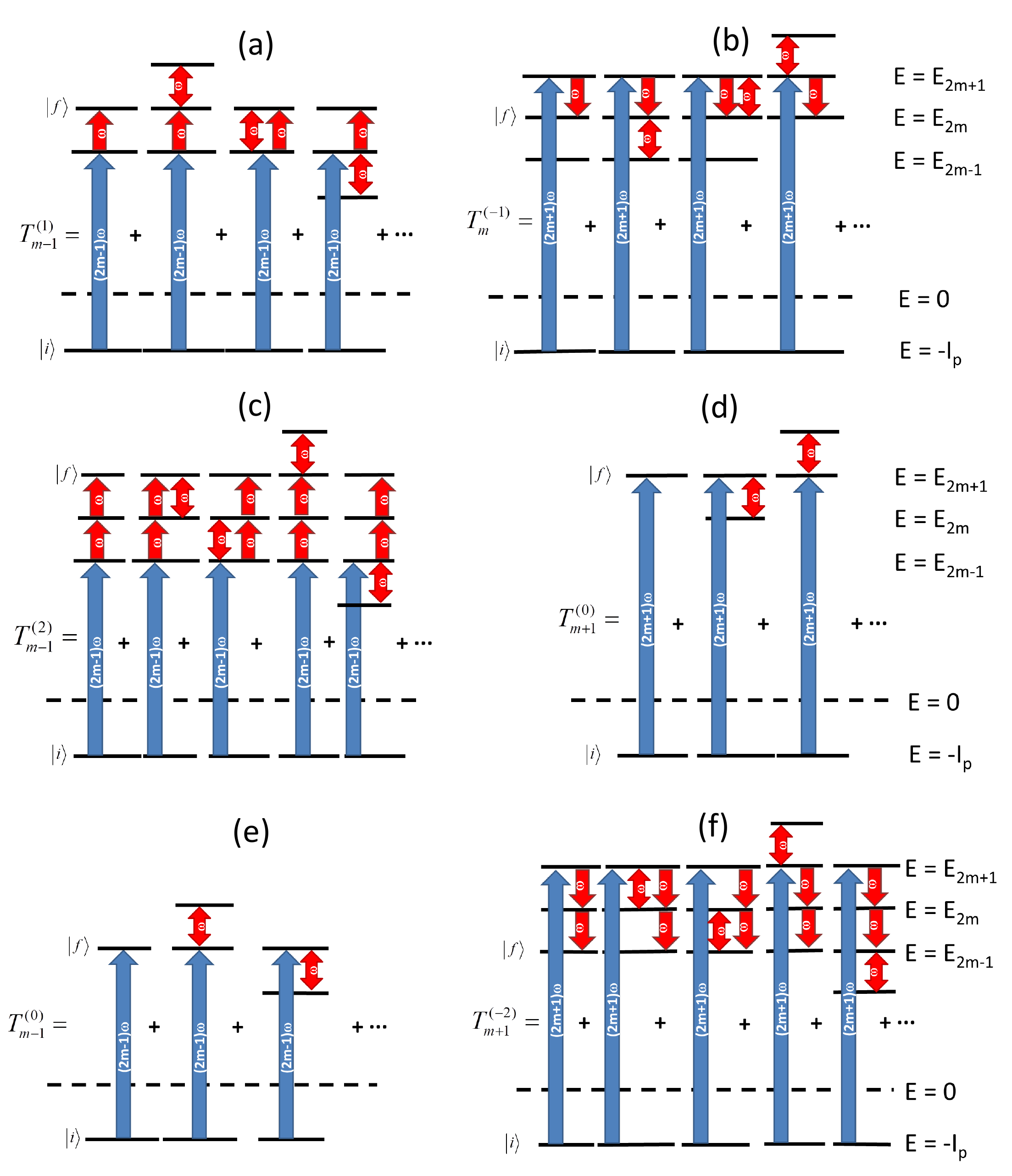}
\centering
\caption{ RABBIT protocol for  interfering pathways from
the initial bound state $\left\vert i\right\rangle $ to final states 
$\left\vert f\right\rangle $ in the continuum.
(a) and (b) show the final ATI state (main SB) involving paths with two or four photons
(absorption of one $(2m-1)\omega$, $(2m+1)\omega$ respectively and absorption-emission of probe photons). (c) and (d) show the final ATI state [HH($15$)] involving
paths with three or five [for (c)] and one or three [for (d)] photons (absorption of one $(2m-1)\omega$, $(2m+1)\omega$ respectively and absorption-emission of probe photons). (e) and (f) show the final ATI state [HH($14$)] involving paths with one or three [for (e)] and three or five [for (f)] photons [absorption of one $(2m-1)\omega$, $(2m+1)\omega$ respectively and absorption-emission of probe photons].}
\label{rabbit diagram}
\end{figure}

\begin{figure}[!htbp]
    \centering \includegraphics[width=0.35\textwidth]{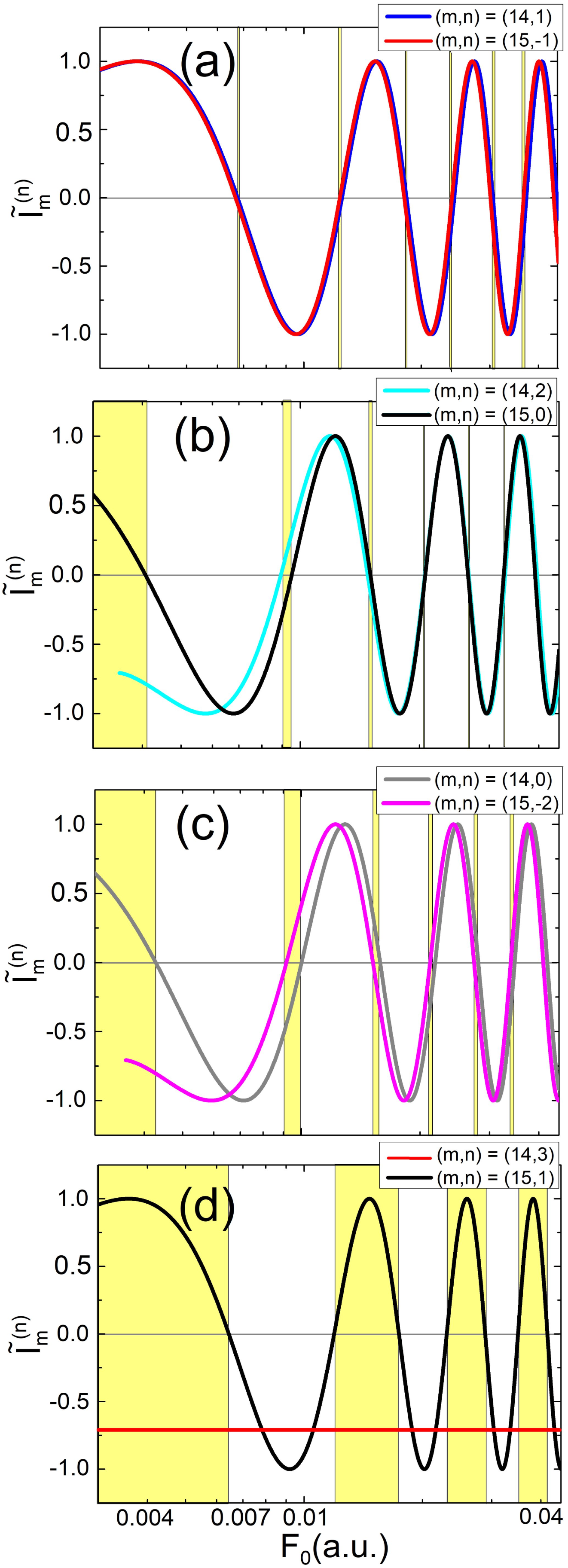}
    \caption{Normalized intracycle factor $\tilde{I}_m^{(n)}$ as a function of the fundamental laser field strength (in logarithmic scale) for energies corresponding to (a) main sideband SB, (b) HH($15$), (c) HH($14$) and (d) one marginal sideband. For each panel we show the intra-cycle factors for paths (a): $\tilde{I}_{14}^{(1)}$ and $\tilde{I}_{15}^{(-1)}$ drawn with blue and red lines, respectively; (b): $\tilde{I}_{14}^{(2)}$ and $\tilde{I}_{15}^{(0)}$ drawn with cyan and black lines, respectively; (c) $\tilde{I}_{14}^{(0)}$ and $\tilde{I}_{15}^{(-2)}$ drawn with gray and magenta lines, respectively; (d) $\tilde{I}_{14}^{(3)}$ and $\tilde{I}_{15}^{(1)}$ drawn with red and black lines, respectively. Yellow (white) regions correspond to anomalous (normal) behavior (see text for explanation).}
    \label{Imn tilde}
\end{figure}

Considering Eq. (\ref{Tm4}), the transition probability [Eq. (\ref{T-RABBIT})] to any 
ATI peak (HH or sideband) can be written as \cite{boll23}
\begin{equation}
\left\vert T\right\vert ^{2} = A+B\cos \left( 2\phi + \delta\right) , \label{T-fit} 
\end{equation}
where $A$ and $B$ depend on each contribution and $\delta$ is the phase delay given by
\begin{equation}
\delta =\underset{\text{{\small HH}}}{\underbrace{\phi_{m}-\phi _{m-1}}}+\underset{\text{{\small {}atomic}}}{\underbrace{\arg \left[ \tilde{I}_{m-1}^{(n_1)}\right] -\arg \left[ \tilde{I}_{m}^{(n_2)}\right] \,}},
\label{delta-SB}
\end{equation}
where the first two terms of the right hand side of Eq. (\ref{delta-SB}) correspond to 
the group delay of each HH($m$) and HH($m-1$) of the XUV field when arriving at the target and the last two terms correspond to the atomic phase delays, which contain the phase inherent to the photoionization and measurement processes (see Appendix \ref{A-SD-HH} for a detailed calculation).
In Eq. (\ref{delta-SB}), $n_1$ and $n_2$ determine the final energy from Eq. (\ref{SB}) and must satisfy the relation $n_1 = n_2+2$.
In Table \ref{Table-AB} (Appendix \ref{A-SD-HH}) we list the corresponding values of the parameters A,B and $\delta$ for the relevant peaks considered here, namely, main SB,  HH($m$), and HH($m-1$), with energies $E_{2m}$, $E_{2m+1}$, and $E_{2m-1}$, respectively.
In our SCM, the population of every ATI peak (sideband or HH) is the result of the interference of two different contributions with a multitude of quantum paths, as depicted in Fig. \ref{rabbit diagram}.
The normalized intracycle factors $\tilde{I}_m^{(n)}$ are related to the $I_m^{(n)}$ via 
Eq. (\ref{I-tilde-1}) and are shown in Fig. \ref{Imn tilde}.

In the following, we analyze the main aspects of the photoelectron dependence on phase and laser field intensity for these sidebands and HH.

\begin{itemize}
    \item SB analysis

We examine the main sideband (SB) at $E_{2m}$ between HH($m-1$) and HH($m$) [see Eq. (\ref{SB})]. In the perturbative regime, one contributing path corresponds to the absorption of one photon of harmonic order $2m-1$ followed by the absorption of one photon of fundamental frequency, i.e., $n=1$, whereas the other path corresponds to the absorption of one photon of harmonic order $2m+1$ followed by the emission of one photon of fundamental frequency, i.e., $n=-1$. See left diagrams of Figs. \ref{rabbit diagram}a and  \ref{rabbit diagram}b involving two-photon transitions.
However, our non-perturbative SCM includes multiple absorption and emission of probe photons for each of the two contributions with the same final sideband energy (see diagram in Fig. \ref{rabbit diagram}a and \ref{rabbit diagram}b where up to four-photon transitions schemes are displayed). 

In Eq. (\ref{T-fit}) we observe a $\pi $-periodic oscillation of the sideband probability as a function of the relative phase $\phi $ on a background signal $A-B$ which is nearly zero provided $F_{m-1}\simeq F_{m}$. Values of $A$, $B$, and $\delta_{\mathrm{SB}}$ are shown in Table \ref{Table-AB} in Appendix \ref{A-SD-HH}.
To obtain the value of the phase delay, we have to analyze carefully the sign of the phase, namely, if $\tilde{I}_{m-1}^{(1)}$ and $\tilde{I}_{m}^{(-1)}$ have equal signs or not (see Appendix \ref{A-SD-HH} for details). We denote as the \textquotedblleft normal\textquotedblright\ behavior: when the two emission ($T_m$) and absorption ($T_{m-1}$) paths have the same sign (see Fig. \ref{Imn tilde}a).
Therefore, when the two neighbor harmonics HH($m$) and HH($m-1$) of the given sideband $2m$ considered in RABBIT have the same phases, i.e., $\phi _{m}=\phi _{m-1}$, we get that the phase delay
\begin{equation}
\delta _{\mathrm{SB}}=0,
\end{equation}
in agreement with the \textquotedblleft rule of thumb\textquotedblright \ proposed in Ref. \cite{Bertolino21}. 
Nevertheless, there are tiny regions in the $F_{0}$ domain for which $\tilde{I}_{m-1}^{(1)}$ and $\tilde{I}_{m}^{(-1)}$ has opposite sign. For these values of $F_{0}$, the ``anomalous'' behavior indicates a sideband phase delay $\delta _{\mathrm{SB}}=\pi$. We believe that a measurement of these anomalous sideband phase delays is a formidable task since an integration over the (probe) laser intensity due to the average over the focal volume would preclude the anomalous phase delays over the normal phase delays since the latter is much more probable (more than two orders of magnitude) than the former. In Fig. \ref{Imn tilde}a we show the imaginary part of the two absorption and emission terms. The normal behavior applies to the $F_0$ or intensity domain except for a few intervals depicted as yellow (gray) very narrow stripes. 

We now pass to analyze the marginal sidebands formed for energies below HH($m-1$) at $E_{2m-1}$ and above HH($m$) at $E_{2m+1}$, which are the result of the interference of two contributions: $T_{m-1}^{(n+2)}$ and 
$T_{m}^{(n)}$ with $n=\cdots,-4,-3,1,2,\cdots $.
Analogously, we find sidebands with a phase delay $\delta=0$
as the \textquotedblleft normal\textquotedblright\ behavior with few
exceptions corresponding to the \textquotedblleft
anomalous\textquotedblright\ behavior of $\delta=\pi$, which would not survive a focal volume average, except for low probe intensities see Fig. \ref{Imn tilde}d.
Thus, marginal sidebands are in opposite phase ($\delta = \pi$) with respect to the main SB ($\delta_{\mathrm{SB}}$ = 0) for weak probe lasers ($I \lesssim 2 \times 10^{12}$ W/cm$^2$).
We want to point out that the effect of the Coulomb potential of the remaining core on the ejected electron, neglected in our SCM, is analyzed in the following section when we compare with the results of the full \textit{ab initio} TDSE.

\item HH analysis

Similarly to the analysis of the sidebands, we can calculate the variation of the population of the HH($m$) at energy $E_{2m+1}$ as a function of the relative phase $\phi$.
In this case, we consider the interference of the following two different probability amplitudes: one corresponding to the absorption of one pump photon HH($m-1$) to $E_{2m-1}$ plus the absorption of at least two photons of fundamental frequency, i.e., 
$T_{m-1}^{(2)}$, and the other corresponding to the direct absorption of only one photon of harmonic of order HH($m$) to $E_{2m+1}$ and an equal number of absorption and emission of probing photons, i.e., $T_{m}^{(0)}$. Schemes of some paths that lead to this energy peak can be found in Figs. \ref{rabbit diagram}c and \ref{rabbit diagram}d. Lowest order perturbation theory for the absorption path corresponds to the left diagram in Fig. \ref{rabbit diagram}c. 
The next order in the perturbative approach involves five-photon transitions as seen in the rest of the diagrams of the same figure. The lowest order perturbation diagrams of the direct path are displayed at the left of Fig. \ref{rabbit diagram}d involving a one-photon transition. The following order involves three-photon transitions (see Fig. \ref{rabbit diagram}d). When $F_{m-1}\simeq F_{m},$ the modulation of the transition amplitude in Eq (\ref{T-fit}) has a positive background $A-B$ (see Table \ref{Table-AB}
in appendix \ref{A-SD-HH}).

In Fig. \ref{Imn tilde}b we observe that in the vast majority of the intensity of the probe (or $F_{0}$) domain $\tilde{I}_{m-1}^{(2)}$ and $\tilde{I}_{m}^{(0)}$ has the same sign and consequently $T_{m-1}^{(2)}$ and $T_{m}^{(0)}$ are in phase (white shaded area). We can call this as the \textquotedblleft normal\textquotedblright\ behavior for which $\delta_{\mathrm{HH}(m)}=0$.
In contrast, there are small regions in the $F_{0}$ domain for which we have the opposite behavior. This \textquotedblleft anomalous\textquotedblright\ behavior is depicted in Fig. \ref{Imn tilde}b in yellow (gray) shade in the intensity (or $F_{0}$) domain.
For these values of $F_{0}$, the anomalous behavior indicates a sideband phase delay $\delta_{\mathrm{HH}}(m)=\pi$. In general, as any measurement corresponds to intensity integration up to a peak value due to the average over the focal volume, the anomalous behavior would not survive this integration due to its much lower ionization probability in comparison to the normal behavior. This is a general rule with one crucial exception: the anomalous behavior extends for intensities lower than $5.9\times 10^{11}$ W/cm$^{2}$ ($F_{0} \le 0.0041$ a.u.).
In the perturbative regime of the probe pulse, the main SB at $E_{2m}$ and the HH($m$) at $E_{2m+1}$ are in phase opposition in agreement with the rule of thumb  \cite{Dahlstrom13,Guenot14,Bertolino21}. However, for probe laser intensities higher than this limit, the main sideband SB and the HH($m$) are in phase.

Analogously, we can consider the interference of the following two different probability amplitudes: $T_{m-1}^{(0)}$ and $T_{m}^{(-2)}$ for the HH($m-1$) analysis.
We observe in Fig. \ref{Imn tilde}c that in the vast majority of the intensity (or $F_{0}$) domain the direct path is in phase with respect to the emission path. Both direct and emission terms will be essentially in phase as observed in Fig. \ref{Imn tilde}c for the \textquotedblleft normal\textquotedblright\ behavior resulting in
$\delta _{\mathrm{HH}(m-1)}=0$.
In contrast, the small regions in the $F_{0}$ domain for which 
$\tilde{I}_{m-1}^{(0)}$ and $\tilde{I}_{m}^{(-2)}$ have different signs corresponds to the \textquotedblleft anomalous\textquotedblright\ behavior as depicted in
Fig. \ref{Imn tilde}c in yellow (gray) shade in the intensity (or $F_{0}$) domain indicating a sideband phase delay $\delta _{\mathrm{HH}(m-1)}=\pi$. As the anomalous behavior extends for intensities lower than $6.5\times 10^{11}$ W/cm$^{2}$ ($F_{0} \le 0.0043$ a.u.), below this value, the sideband at $E_{2m}$ and the HH($m-1$) at $E_{2m-1}$ are in phase opposition in agreement with the rule of thumb developed in the perturbative regime of the probe pulse \cite{Dahlstrom13,Guenot14,Bertolino21}. 
However, for moderately strong probe lasers the main sideband SB and the harmonic 
HH($m-1$) are in phase.

\end{itemize}

\begin{figure}[!htbp]
\centering
\includegraphics[width=0.9\textwidth]{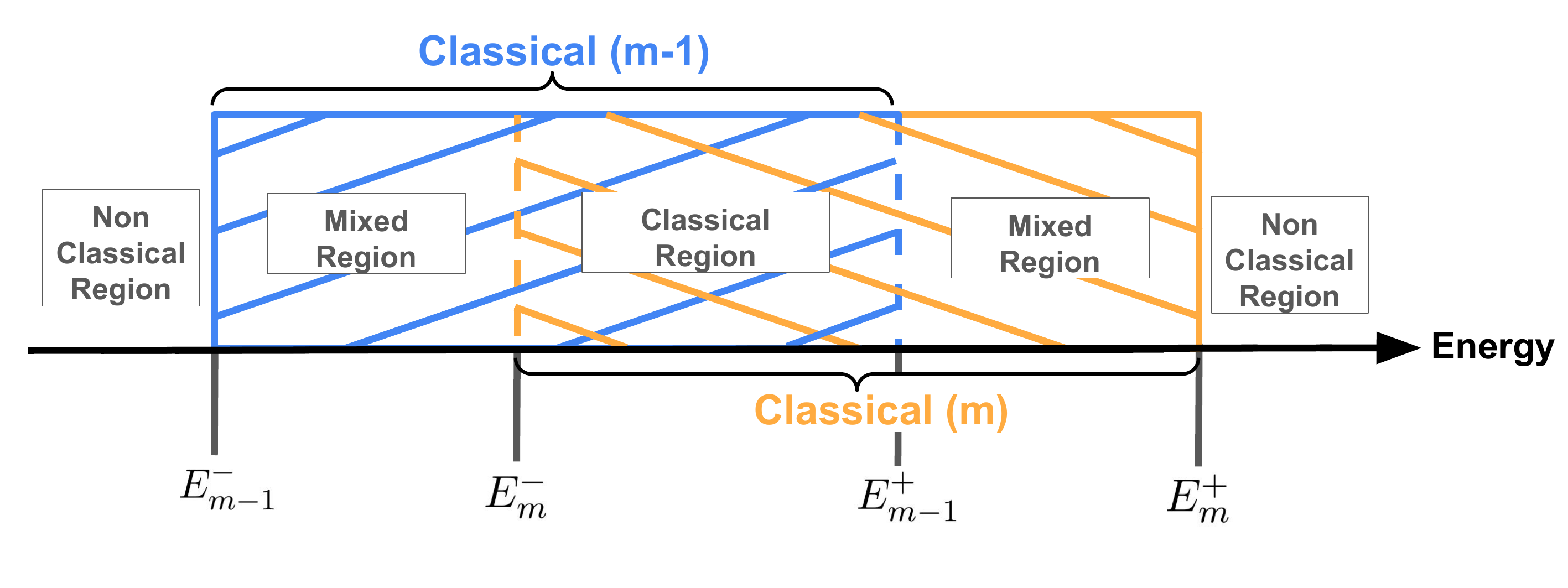}
\caption{Energy diagram where the limits of each region are specified. The classical region is such that the energy lies between $E_{m}^-$ and $E_{m-1}^+$ (region with orange and blue stripes). The mixed regions go between $E_{m-1}^-$ and $ E_{m}^-$ (region with only blue stripes) and between $E_{m-1}^+$ and $E_{m}^+$ (region with only orange stripes).
The non-classical regions (region without stripes) correspond to energies lower than 
$E_{ m-1}^-$ or higher than $E_{m}^+$.
$E_{ m}^{-+}$ and $E_{ m-1}^{-+}$ are the classical limits of $E_{ 2m+1}$ and $E_{2m-1}$, respectively, with $E_{ m-1}^-$ = $(v_{0,m-1} - \frac{F_0}{\omega})^2/2$, $E_{ m}^-$ = $(v_{0,m} - \frac{F_0}{\omega})^2/2$, $E_{ m-1}^+$ = $(v_{0,m-1} + \frac{F_0}{\omega})^2/2$ and $E_{ m}^+$ = $(v_{0,m} + \frac{F_0}{\omega})^2/2$.  }
\label{complex-plots-limits}
\end{figure}

We have already analyzed that when 
$\left\vert \beta_{+,m}^{(n)}\right\vert <F_{0}/\omega$ two real ionization times arise demarcating the classically allowed region (Fig. \ref{complex-plots-limits}).
The classical limits are given by the transcendental equation $\left\vert \beta _{+,m}^{(n)}\right\vert=F_{0}/\omega ,$ which can be numerically solved (see Appendix \ref{Regions} for detailed calculations). 
Now, we extend our SCM to non-classical allowed regions in the $F_0$ domain $\bigg(\left\vert \beta_{+,m}^{(n)}\right\vert > F_{0}/\omega\bigg)$ with complex times 
[Eq. (\ref{stcomplex-2})].
Each of the contributions $T_m$ and $T_{m-1}$ in Eq. (\ref{T-RABBIT})
has a classical allowed region in the energy domain $((v_{0,m}-F_0/\omega)^2,(v_{0,m}+F_0/\omega)^2)$ and $((v_{0,m-1}-F_0/\omega)^2,(v_{0,m-1}+F_0/\omega)^2)$, respectively, surrounded by classically forbidden regions (non-classical and mixed Region) as exhibited in the schematic Fig. \ref{complex-plots-limits}.
In the mixed region, there are certain
regions in the $F_{0}$ domain where the phase delay is $\delta = 0$ and other regions where $\delta = \pi$. In the non-classical regions, $\delta = \pi$ for the whole $F_{0}$ domain. For more details, see Appendix \ref{Regions}.

\section{Results}
\label{results}

\begin{figure}[!htbp]
    \centering \includegraphics[width=0.9 \textwidth]{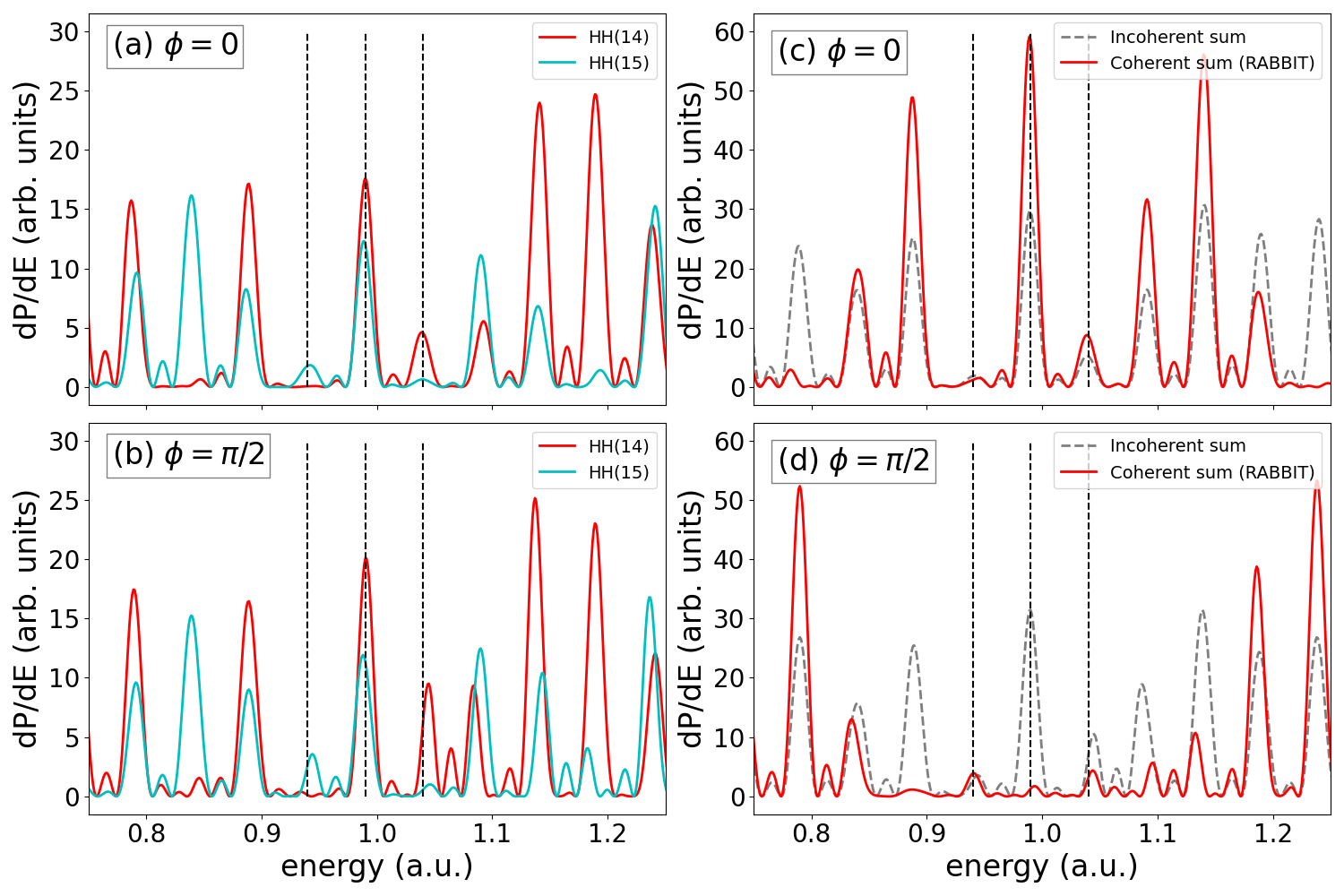}
    \caption{Photoelectron spectrum for forward emission of hydrogen as a function of the emission energy calculated within the SFA theory. The HH frequencies correspond to overtones 29 and 31 ($m = 14, 15$) of fundamental frequency $\omega = 0.05$ a.u.. 
    (a) and (b): one-frequency LAPE processes HH($14$) in red solid line and HH($15$) in cyan line for phases $\phi = 0$ and $\phi = \pi/2$, respectively. 
    (c) and (d): Coherent sum for RABBIT $\vert T_{14} + T_{15}\vert^2$ for phases $\phi = 0$ and $\phi = \pi/2$, respectively, contrasted with the incoherent sum 
    $\vert T_{14}\vert^2 + \vert T_{15}\vert^2$ in dashed line, for phases $\phi = 0$ and $\phi = \pi/2$, respectively.
    Field intensities are $F_0 = 0.01$ a.u. and $F_{14,15} = 0.01$ a.u.. Calculations are made for flattop envelopes with linear one-cycle ramps on and off.
    The attosecond pulse train of XUV duration comprises 3 cycles of the fundamental component. Vertical dashed lines correspond to energies of the HH($14$), main SB, and HH($15$) peaks.}
    \label{Fig_coeherent_incoherent}
\end{figure}

In this section, we refer to the SFA as the numerical solution of the transition
amplitude in Eq. (\ref{Tif}) when the final channel is represented by the
Volkov wave function in Eq. (\ref{Volkov}) within the dipole approximation
in the length gauge. Similarly, we refer to TDSE as the numerical solution of
Eq. (\ref{TDSE}) \cite{Tong00,Arbo06b,Arbo08a}. In the following,
we compare the results for photoionization of atomic hydrogen calculated within the SCM to the counterpart obtained using the SFA and TDSE.
Calculations are made for flattop envelopes with linear one-cycle ramps on and off with probe laser frequency $\omega = 0.05$ a.u.. Field intensities are $F_0 = 0.01$ a.u. and $F_m = 0.01$ a.u. for $m=14$ and $15$. The attosecond pulse train of XUV duration comprises 3 cycles of the fundamental component.
First, we show the (lack of) dependence of the forward photoelectron spectra on the relative phase $\phi$ between the probe field and the pump pulse train in two isolated LAPE processes with the pump of only one high frequency HH($14$) or HH($15$) with corresponding emission probabilities $\vert T_{14}\vert^2$ or $\vert T_{15}\vert^2$, respectively.
In Fig. \ref{Fig_coeherent_incoherent}a and \ref{Fig_coeherent_incoherent}b, we display the SFA photoelectron spectrum in the forward direction for two different NIR phases, $\phi=0$ and $\pi/2$, respectively. It can be noted that the NIR phase variations due to the presence of the envelope (ramps) of the short pulse are very small, justifying its neglect in the SCM [Eq. (\ref{LAPE})] \cite{Gramajo16,Gramajo18}. 
In contrast, in Fig. \ref{Fig_coeherent_incoherent}c and \ref{Fig_coeherent_incoherent}d we show the energy distribution in the forward direction for the RABBIT-like protocol,
where the pump pulse is comprised of two harmonics HH($14$) and HH($15$), calculated within the SFA with the same laser parameters.
We see that the coherent sum of the two contributions $\vert T_{14}+T_{15}\vert^2$
is considerably phase-dependent with a strong enhancement due to constructive interference in the region around the main sideband (SB) at $E_{30}=0.9984$ a.u. [Eq. (\ref{SB})] for $\phi=0$, while we observe depletion due to destructive interference of the main SB for $\phi=\pi/2$. We contrast the result with the incoherent sum of the two individual contributions, $\vert T_{14} \vert^2+\vert T_{15} \vert^2$ (dashed line).

\begin{figure}[!htbp]
\centering
\includegraphics[width=0.7 \textwidth]{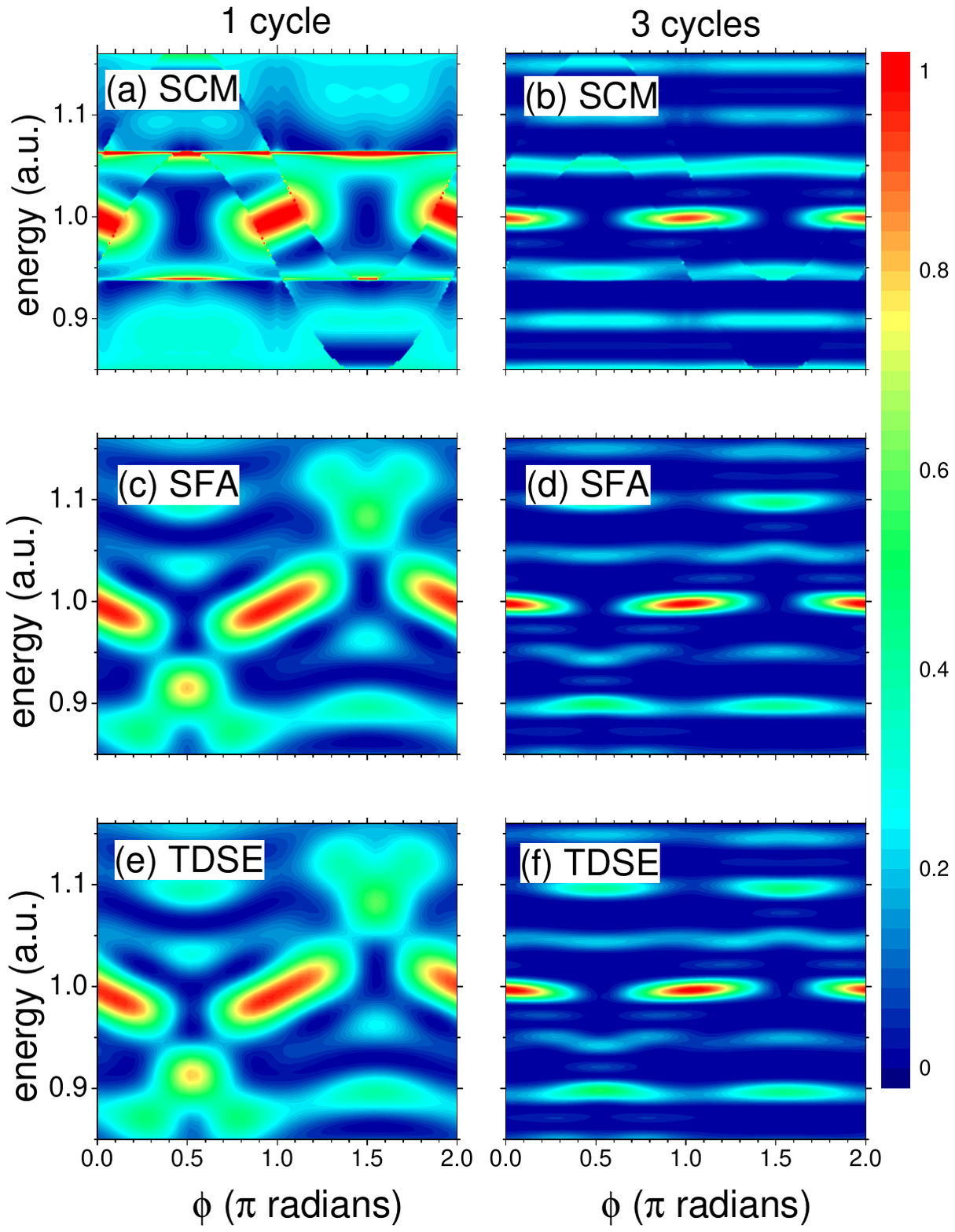}
\caption{Photoelectron spectra for ionization of hydrogen in the forward direction as a function of the emission energy and the relative phase $\phi$ between the attosecond pulse train and the probing laser. Photoelectron spectra are calculated within the SCM [(a) and (b)], SFA [(c) and (d)], and TDSE [(e) and (f)] for the XUV duration of one optical cycle of the probe laser in panels (a), (c), and (d) and three optical cycles of the probe laser in panels (b), (d), and (f). The laser field strength is $F_0 = 0.004$ a.u. and all distributions are normalized to unity.}
\label{intrainter-p004}
\end{figure}

For a thorough analysis of the electron yield in Fig. \ref{intrainter-p004} we plot
the energy photoelectron spectrum as a function of the relative phase $\phi$ for the
RABBIT protocol. The probe NIR pulse has a strength $F_0 = 0.004$ a.u. and frequency $\omega = 0.05$ a.u.. We use different durations of the pump XUV pulse composed of the two harmonics HH(14) and HH(15): one optical cycle of the probe (left column) and three optical cycles of the probe (right column). The main SB is at $E_{30}=0.9984$ a.u.,
the HH(14) at $E_{29}=0.9484$ a.u., and the HH(15) at $E_{31}=1.0484$ a.u.. 
The SCM, SFA, and TDSE intracycle interference patterns of Figs. \ref{intrainter-p004}a, \ref{intrainter-p004}c, and \ref{intrainter-p004}e, respectively, are $2\pi$ periodic, whereas the interplay of the inter- and intracycle interference renders the distribution
rather $\pi$ periodic in Figs. \ref{intrainter-p004}b, \ref{intrainter-p004}d, and \ref{intrainter-p004}f, respectively. In Fig. \ref{intrainter-p004}a the $\phi$-dependent oscillating discontinuity of the $\phi$-dependent photoelectron spectrum is clearly observed, as described in Sec. \ref{theory}. As expected, this discontinuity disappears in the quantum SFA and TDSE distributions due to the width of the wave packet (Figs. \ref{intrainter-p004}c and \ref{intrainter-p004}d).
For the case of XUV pulse duration of three optical cycles, the main SB maximizes at $\phi=0$ and $\pi$, corresponding to $\delta_{\mathrm{SB}}=0$, whereas HH maximizes at $\pi=\pi/2$ and $3\pi/2$ corresponding to 
$\delta_{\mathrm{HH}(14)}=\delta_{\mathrm{HH}(15)}=\pi$ [see Eq. (\ref{T-fit})].
This is in agreement with the ``rule of thumb'' calculated in the perturbative regime \cite{Guenot12,Guenot14,Dahlstrom13,Bertolino21}.
In addition, Fig. \ref{intrainter-p004}a shows the characteristic caustics of SCM at $E_{15}^- = 0.937$ a.u. and $E_{14}^+ = 1.063$ a.u. defined in appendix \ref{A-SD-HH} and 
Fig. \ref{complex-plots-limits}. Finally, the marginal sidebands with energies below HH($14$) and above HH($15$) have the same phase delays as the harmonics.

\begin{figure}[!htbp]
\centering
\includegraphics[width=0.7 \textwidth]{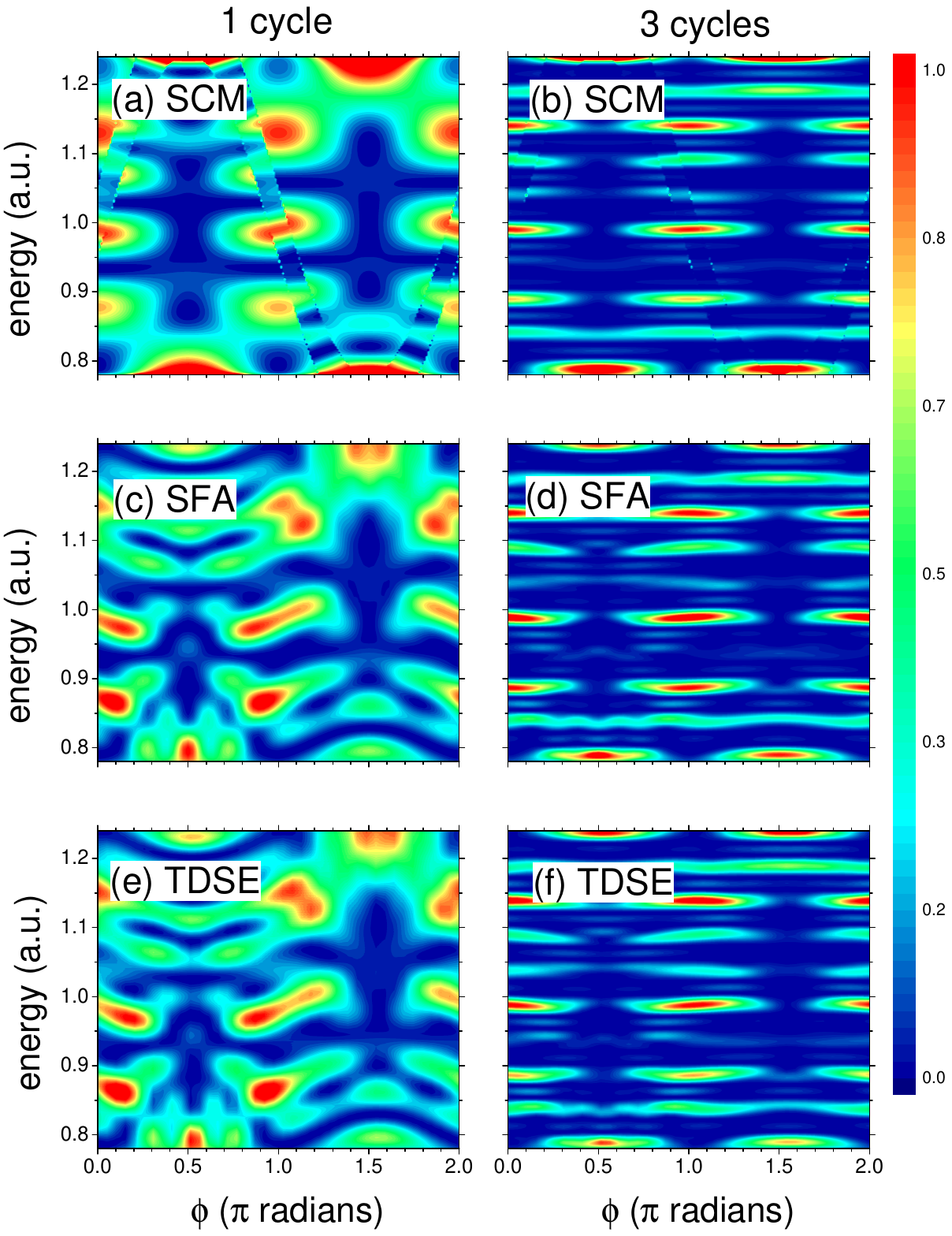}
\caption{The same photoelectron spectra as in Fig \ref{intrainter-p004}, but with field strenght $F_0 = 0.01$ a.u..}
\label{intrainter-p01}
\end{figure}

When we increase the NIR laser field strength to $F_0 = 0.01$ a.u., the main SB, HH($14$), and HH($15$) locate at $E_{30} = 0.99$ a.u., $E_{29} = 0.94$ a.u., and $E_{31} = 1.04$ a.u., respectively, and the classical allowed region corresponds to
$E \in (E_{15}^- = 0.78$ a.u. $, E_{14}^+ = 1.246$ a.u.$)$.
In Fig. \ref{intrainter-p01} we observe that the main sideband, HH(14), HH(15),
and marginal sidebands are all in phase within the classical region with phase delays
$\delta_{\mathrm{HH}(14)} = \delta_{\mathrm{HH}(15)} = \delta_{\mathrm{SB}} = 0$,
whereas the marginal sidebands in the mixed region (see Fig. \ref{complex-plots-limits}) hold phase delays $\delta = \pi$, as predicted by the SCM. In the photoelectron spectra of Figs. \ref{intrainter-p01}a and \ref{intrainter-p01}b calculated within the SCM, the $\phi$-oscillating discontinuities are visible, however, discontinuities disappear in SFA and TDSE calculations, as expected. The resemblance between the three models, especially SFA in Figs. \ref{intrainter-p01}c and \ref{intrainter-p01}d and TDSE in Figs. \ref{intrainter-p01}e and \ref{intrainter-p01}f is outstanding, showing that the
action of the Coulomb potential of the remaining core on the wavepacket of the emitted electron is lower than the numerical error in our TDSE calculations for the energy and laser parameters considered, i.e., about $0.16$ rad for phase delays or $40$ as for time delays. It can be noted that for these high emission energies our SCM describes qualitatively the interference features of the spectra, which strongly supports the time-domain interference viewpoint of the photoionization process.

\begin{figure}[!htbp]
\centering
\includegraphics[width=0.7 \textwidth]{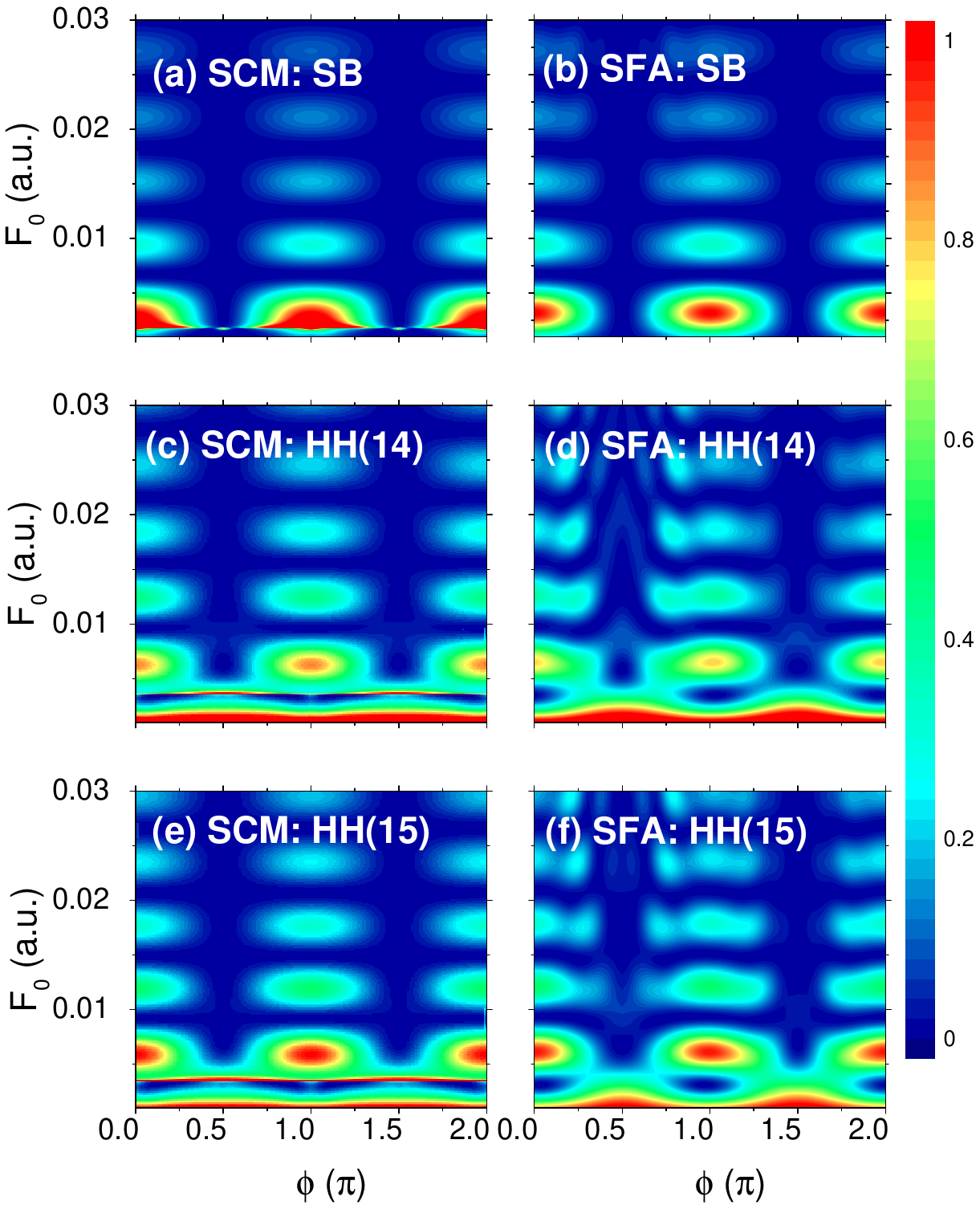}
\caption{Forward emission photoelectron spectra as a function of the NIR laser peak field $F_0$ and phase $\phi$ calculated within the SCM (left column) and SFA (right column). Emission energies correspond to SB in pannels (a) and (b), HH($14$) in pannels (c) and (d), and HH($15$) in panels (e) and (f). Pulse parameters and SFA envelope are the same as Fig. \ref{Fig_coeherent_incoherent}. }
\label{Fig_F0_scan}
\end{figure}

\begin{figure}[!htbp]
\centering
\includegraphics[width=0.7 \textwidth]{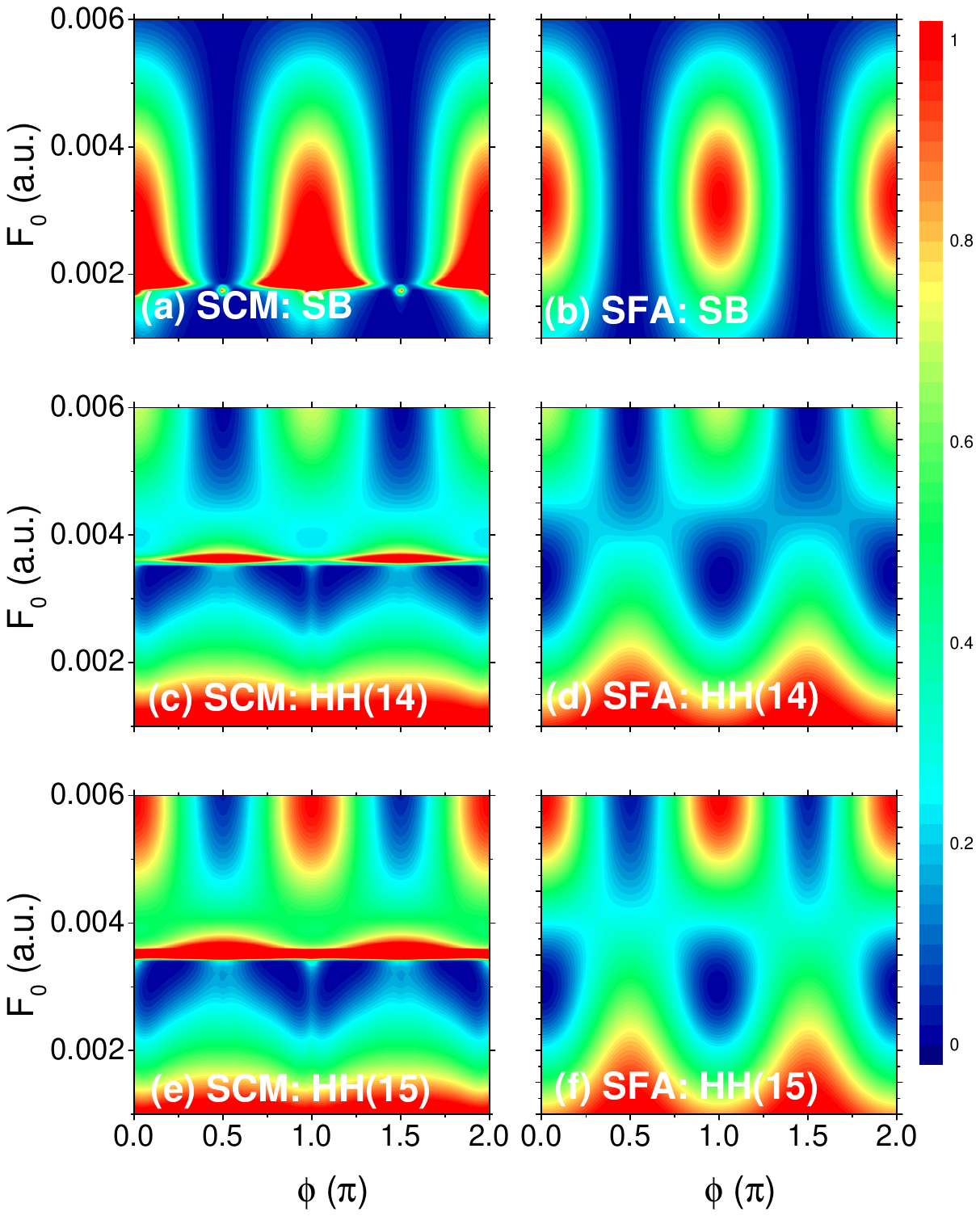}
\caption{Close up of Fig. \ref{Fig_F0_scan} enhancing the low intensity probe field. }
\label{Fig_F0_scan-zoom}
\end{figure}

To get more insight about the behavior of the phase delays as a function of the NIR laser intensity, we show in Fig \ref{Fig_F0_scan} the photoelectron spectra of hydrogen for electron emission in the forward direction for the three energies peaks: HH($14$), main SB, and HH($15$). We observe that, except for low values of $F_0$, distributions maximize at $\phi = 0$ and $\pi$ showing that harmonics HH($14$) and HH($15$) and the
main SB are all in phase with
$\delta_{\mathrm{HH}(14)} = \delta_{\mathrm{HH}(15)} = \delta_{\mathrm{SB}} = 0$.
In Fig. \ref{Fig_F0_scan-zoom} we observe a close-up of Fig. \ref{Fig_F0_scan} for low intensities of the probe field.
We need to remark that the SCM predicts all the main features shown by the SFA, even for small values of NIR laser strength $F_0$.
The SCM exhibits the caustics characteristic of any semiclassical theory \cite{Barrachina89, Kelvich17} at $F_0 \simeq 0.0037$ a.u. for the HH($14$) and HH($15$), and the main SB at $F_0 \simeq 0.0018$ a.u., which are an artifact solvable when using more sophisticated theories like the uniform saddle-point approximation \cite{Faria02}. However, our aim is not providing an accurate model since we already count with the SFA and TDSE calculations, but shedding some light on the time-dependent electron processes involved in RABBIT-like setups beyond the perturbative regime of low probe intensities. We observe that the general trend of the SFA is also reproduced by the SCM, despite the mentioned caustics. Firstly, in Fig. \ref{Fig_F0_scan-zoom} we observe that for low probe intensities the SCM and SFA distributions maximize at 
$\phi = 0$ and $\pi$ for the main sideband SB and $\phi = \pi/2$ and $3\pi/2$ for the HH($14$) and HH($15$)
with ensuing $\delta_{\mathrm{SB}} = 0$ and $\delta_{\mathrm{HH}} = \pi$, respectively,
as stated by the rule of thumb \cite{Dahlstrom12,Dahlstrom13,Bertolino21}. The maxima of
the main SB are essentially invariant with the probe laser intensity.
In contrast, HH($14$) and HH($15$) show a transition from maxima at $\phi = \pi/2$ and $3\pi/2$ for low intensities (corresponding to $\delta_{\mathrm{HH}} = \pi$) to maxima at $\phi = 0$ and $\pi$ for high intensities (corresponding to 
$\delta_{\mathrm{HH}} = 0$) at a defined probe strength $F_0 \simeq 0.0042$ a.u. corresponding to $I \simeq 6 \times 10^{11}$W/cm$^2$.
In principle, the SCM is only valid in the classical region of Fig. \ref{complex-plots-limits}, however, a stretching of the SCM to the mixed and non-classical region provides sensible outcomes comparable to the SFA and TDSE, as observed in Fig. \ref{Fig_F0_scan-zoom} for $F_0$ values lower than the caustics.

\section{Conclusions}
\label{conclusions}

We have developed a semiclassical strong-field theory based on the saddle-point approximation for the atomic ionization by a linearly polarized laser pulse train formed by two neighbor odd harmonics of a concomitant laser of fundamental frequency, responsible for continuum-continuum transitions.
Very importantly, our model includes multi-photon transitions beyond the common approaches extensively studied in the literature, which are based on perturbation treatments of the probe pulse (RABBIT).
As far as we know, this is the first detailed time-dependent analysis of the phase delays in the RABBIT protocol beyond the perturbative regime of the probe laser pulse.
We have studied the photoionization emission of atomic hydrogen from the ground state along the polarization direction of the laser fields.
We have found that there are different types of behavior of the quantum interference processes of the photoelectron spectra from which the phase delays are extracted.
For weak probe laser pulses, the interference patterns of the harmonics and the main sideband between them are in opposition of phase as a function of $\phi$, being the phase delay zero for the main sideband energy and $\pi$ for the two harmonic energy peaks, in agreement with the RABBIT protocole in the perturbative regime of the probe pulse.
In turn, the extension of our study to stronger probe lasers in a RABBIT-like protocole reveals novel effects:
For intensities higher than a threshold about $~5 \times 10^{11}$ W/cm$^2$, 
both, the main sideband and the harmonic peaks stay in phase, with ensuing zero delay.
Our non-perturbative SCM permits us to shed light on the origin of the interferences involved on the photoionization process. In conclusion, we prove that the intracycle interference, corresponding to electron wave packets released within the same NIR optical cycle, rules the behavior of the relative-phase dependent photoelectron spectra and, thus, the phase delays.
By comparing our SCM to the numerical solutions of the SFA and TDSE we conclude that
(i) our analysis is reliable and useful for understanding the physics of the interferometric RABBIT-like method beyond the perturbative treatment of the probe pulse,
and (ii) the time delays due to the action of the Coulomb potential of the remaining core on the wavepacket of the emitted electron is approximately equal or less than the numerical error in our TDSE calculations for the energy and laser parameters considered
(about $40$ as). 
We believe that the present non-perturbative analysis can be very helpful at the time of understanding previous and designing future interferometric RABBIT experiments. 
Therefore, we can conclude on the necessity to revise the extraction method of phase delays beyond the perturbative regime of the probe field.
The present non-perturbative analysis could pave the way for the comprehension of the extraction of phase and time delays through RABBIT-like experiments beyond the perturbative regime of the probe pulse.

\appendix
\section{Ionization times}
\label{Saddle Times Appendix}

We can distinguish two different situations depending on whether there are (or not) real solutions of the saddle equation (\ref{saddle-s}). If 
$\left\vert \beta _{+}(k)\right\vert < F_{0}/\omega$, 
there can be at most two real times that fulfill Eq. (\ref{saddle-s}),
i.e., $t^{(1)}$ and $t^{(2)}$. 
It is important to recall that in the proximity of the coalescence of real times $t^{(1)}$ and $t^{(2)}$ 
[in Eq. (\ref{saddle-s})], this approach is not valid as caustics are present in the transition probabilities \cite{Barrachina89, Kelvich17}.
To avoid caustics we should consider the uniform saddle-point approximation \cite{Faria02}, however, we restrict to the saddle-point approximation to discuss photoionization for the sake of simplicity, since it is not the aim of this work to discuss energies very close to the classical borders.
In turn, if 
$\left\vert \beta _{+}(k)\right\vert >F_{0}/\omega$,
there are no real solutions for the saddle equation (\ref{saddle-s})
and we can say that this region of the momentum map is out of the classical
allowed set of $k$ values. 

The solution of the saddle equation [Eq. (\ref{saddle-s})] with $\beta _{+}(%
\vec{k})<0$ lying in the first half cycle are 
\begin{align}
    \omega t^{(1)}&=\mod{\left( \sin ^{-1}{\left\vert \frac{\omega }{F_{0}}\beta _{+}(\vec{k}%
)\right\vert } - \phi , 2\pi \right)} \nonumber \\
    \omega t^{(2)}&=\mod{\left(\pi - \phi -\sin ^{-1}{\left\vert \frac{\omega }{F_{0}}\beta _{+}(\vec{k}%
)\right\vert } , 2\pi \right) }.
\label{t1}
\end{align}
Instead, if $\beta _{+}(\vec{k})>0$, they lie in the second half cycle and read
\begin{align}
    \omega t^{(1)}& =\mod{\left(\pi +\sin ^{-1}{\ \left\vert \frac{\omega }{F_{0}}\beta _{+}(\vec{k})\right\vert } - \phi , 2\pi \right)} \nonumber \\
    \omega t^{(2)}&=\mod{\left(2\pi -\phi -\sin ^{-1}{\left\vert \frac{\omega}{F_{0}}\beta _{+}(\vec{k}) \right\vert }, 2\pi \right) }. 
    \label{t1p}
\end{align}
This change produces a discontinuity in the photoelectron spectra already discussed in Ref. \cite{Gramajo16}. The modulo function assures that the saddle times lie in the unit cell $t \in \left[0,2\pi \right]$ as $\phi$ varies.

For a more complete analysis of the phase delays across the entire energy range,
we also examine the classical forbidden regions where 
$\left\vert \beta _{+}(k)\right\vert > F_{0}/\omega$ using the saddle-point approximation. The complex times arising from Eq. (\ref{saddle-s}) are
\begin{equation}
\omega t_s= \left\{ \begin{array}{lcc} \mod( \frac{\pi}{2}+  i \cosh^{-1}\left\vert \frac{\omega}{F_0}\beta _{+}(\vec{k})\right\vert - \phi, 2\pi) & &  \beta _{+}(\vec{k}) < -F_0/\omega \\ \\ 
\mod( \frac{3\pi}{2}+  i \cosh^{-1}\left\vert \frac{\omega}{F_0} \beta _{+}(\vec{k})\right\vert - \phi, 2\pi) &  & \beta _{+}(\vec{k}) > F_0/\omega
 \\ \end{array} \right. ,
\label{stcomplex-2}
\end{equation}
where we have used the definition of the vector potential of the NIR laser given by Eq. (\ref{Avector}b).
From Eq. (\ref{stcomplex-2}) we can observe that in the classically forbidden region, the real part of saddle times remains constant. Besides, the imaginary part is independent of $\phi$ and increases rapidly with $\left\vert \beta _{+}(\vec{k})\right\vert$. We point out that the complex conjugate of $t_s$ in Eq. (\ref{stcomplex-2}) must be discarded since the imaginary part of it is negative and it would imply non-physical exponentially growing probabilities.

\section{Calculation of the SB and HH phase delays}
\label{A-SD-HH}

\begin{figure}[!htbp]
\centering \includegraphics[width=0.55\textwidth]{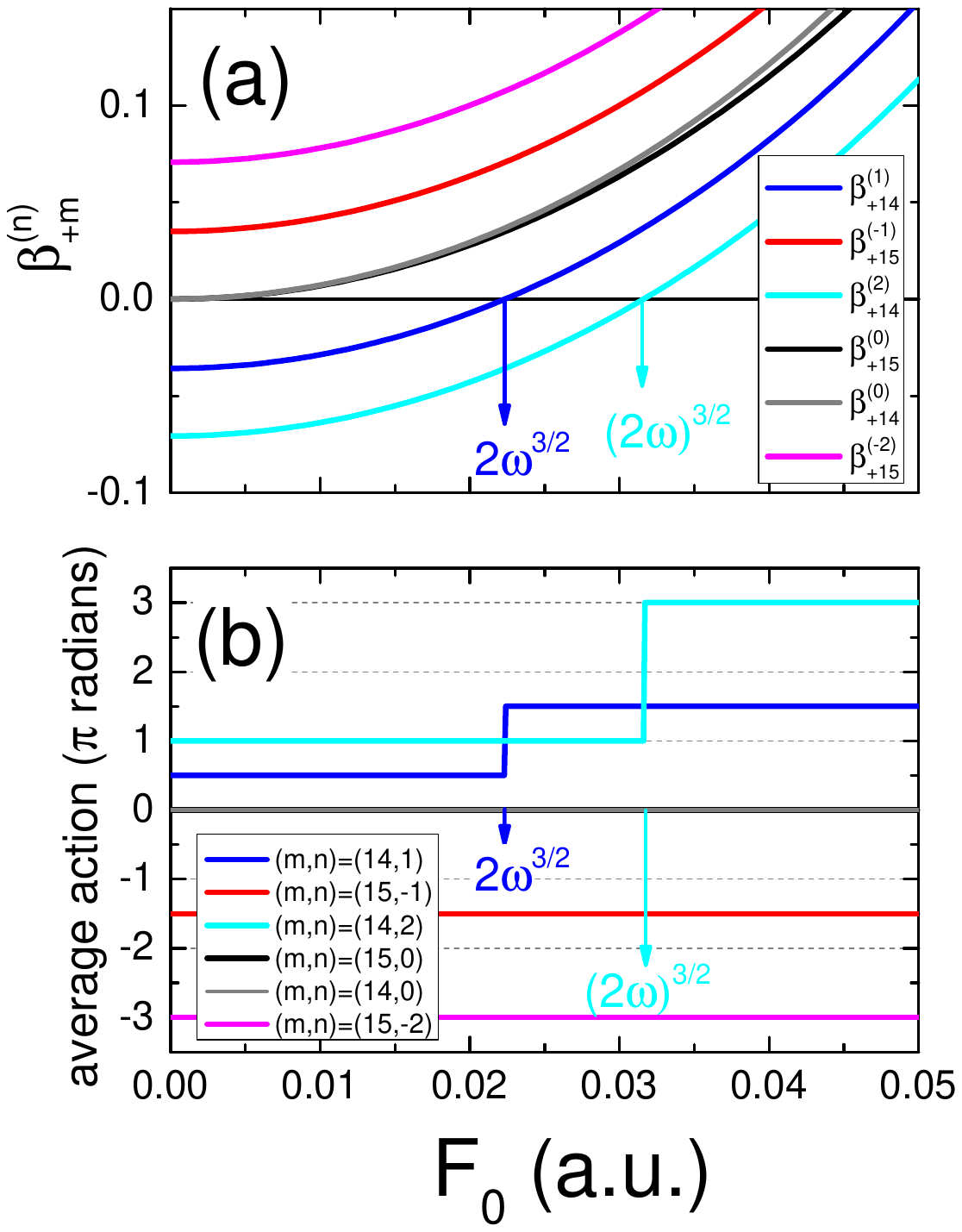}
\caption{(a) $\beta _{+,m}^{(n)}$ as a function of $F_0$ (probe field strength), for the values of $(m,n)$=$(14,1)$; $(15,-1)$; $(14,2)$; $(15,0)$; $(14,0)$; and $(15,-2)$ for blue, red, cyan, black, gray, and magenta, respectively, where the values $2\omega^{3/2}$ and $(2\omega)^{3/2}$ corresponding to the change of  $\beta _{+,m}^{(n)}$ sign for those values of $(m,n)$ are indicated. (b) Average action $\bar{S}_m$ as a function of $F_0$.} \label{beta Savg}
\end{figure}

To determine the sign of $\beta_{+,m}^{(n)}$, from the saddle equation 
[Eq. (\ref{saddle-s})], we can multiply and divide $\beta
_{+,m}^{(n)}$ by the quantity $v_{0,m}+k$, which is clearly positive for
forward emission. Then,
\begin{equation}
\beta _{+,m}^{(n)} = \frac{\left( v_{0,m}-k\right) \left( v_{0,m}+k\right) 
}{v_{0,m}+k}=\frac{v_{0,m}^{2}-k^{2}}{v_{0,m}+k} . 
\label{betanm1}
\end{equation}
The numerator of Eq. (\ref{betanm1}) can be written as
\begin{eqnarray}
v_{0,m}^{2}-k^{2} &=& \left[ \left( 2m+1\right) \omega -I_{p}\right] -\left[ n\omega
+\left( 2m+1\right) \omega -I_{p}-U_{p}\right]  \notag \\
&=& U_{p}-n\omega ,  \label{betanm2}
\end{eqnarray}
where we have used that $k^{2}=2E_{2m+1+n}$ and Eq. (\ref{SB}). Eq. (%
\ref{betanm2}) shows that the sign of $\beta _{+,m}^{(n)}$ does not depend on
the pump harmonic number $m$. For emission paths ($n<0$) we see that $\beta
_{+,m}^{(n)}>0$. However, for absorption paths ($n>0$), $\beta _{+,m}^{(n)}$
can be either positive, negative, or zero, depending on $U_p$ and, thus, the probe peak field. In Fig. \ref{beta Savg}a, we show that for weak fields, i.e., $U_{p}<n\omega$,
or equivalently, $F_{0}<2n^{1/2}\omega ^{3/2}$, the magnitude $\beta_{+,m}^{(n)}<0$
for $n > 0$.
In turn, for $F_{0}>2n^{1/2}\omega ^{3/2}$, the magnitude $\beta_{+,m}^{(n)}>0$
for $n > 0$.
In Fig. \ref{beta Savg}b we also plot the average actions $\bar{S}_m$ 
[defined above Eq. (\ref{ImkSP})] for each different ($m,n$) contributions.
Whereas $\bar{S}_m$ corresponding to net absorption of probe photons ($n>0)$
is defined as piecewise functions of the peak field $F_{0}$,
$\bar{S}_m$ corresponding to emission of probe photons ($n<0$) or direct transitions
with no net exchange of probe photons ($n=0$) is constant [see Eq. (\ref{betanm2})].

In order to examine the phase delay for a particular energy peak, we have to calculate the coherent superposition of two transition amplitudes [Eqs. (\ref{Tsaddle}) and (\ref{Tm4})] as
\begin{eqnarray}
\left\vert T_{m-1}^{(n_1)}+T_{m}^{(n_2)}\right\vert ^{2} &=& N^{2}\left[ F_{m-1}^{2}\left\vert I_{m-1}^{(n_1)}\right\vert^{2} + F_{m}^{2}\left\vert I_{m}^{(n_2)}\right\vert ^{2}+ \right. \label{T-appendix-1} \\
&&\left. 2F_{m-1}F_{m}\left\vert I_{m-1}^{(n_1)}\right\vert
\left\vert I_{m}^{(n_2)}\,\right\vert \cos \left( \phi
_{m-1} - \phi _{m}+\arg \left[ I_{m}^{(n_2)}\right]-\arg \left[
I_{m-1}^{(n_1)}\right] \right) \right], \notag 
\label{T-appendix-2}
\end{eqnarray}
where $n_1$ and $n_2$ determine the final energy [Eq. (\ref{SB})] and must satisfy the relation $n_1 = n_2+2$. Clearly, to inspect the phase delay $\delta$ in Eq (\ref{T-fit}) we have to analyze carefully the argument of the cosine function of Eq. (\ref{T-appendix-1}), particularly $\arg \left[ I_{m}^{(n_2)}\right] $ and $ \arg \left[I_{m-1}^{(n_1)}\right]$. The argument of $I_{m}^{(n)}$ is given by the average action $\bar{S}_m$ that can be calculated by replacing the saddle times given by equations (\ref{t1}) and (\ref{t1p}) into Eq. (\ref{action2}) regarding that in Eq. (\ref{abc}a), $a = n \omega$ for any ATI
energy peak. Therefore, the average action reduces to $\bar{S}_m=- n \phi + n f_m(F_0)$, where $f_m(F_0)$ is a constant piecewise function equal to multiples of $\pi/2$ in the $F_0$ domain, as discussed above. Now, it is convenient for the analysis to express Eq. (\ref{ImkSP}) as
\begin{equation}
I_{m}^{(n)} = 2 g_m(\vec{k}, t^{(\alpha)}) e^{-in\phi} \Tilde{I}_{m}^{(n)}, 
\label{I-tilde-1}
\end{equation}
with
\begin{equation}
\tilde{I}_{m}^{(n)} = e^{i n f_m (F_0)} \cos{\left(\frac{\Delta S}{2}-\frac{\pi}{4} \textrm{sgn}[\beta_+(\vec{k})]\right)}. 
\label{I-tilde-2}
\end{equation}
Finally, introducing Eqs. (\ref{I-tilde-1}) and (\ref{I-tilde-2}) into the argument of the cosine function of Eq. (\ref{T-appendix-1}) and comparing with Eq. (\ref{T-fit}), we find the simple expression for the delay given in Eq. (\ref{delta-SB}).
The corresponding values of the parameters of Eq. (\ref{T-fit}) can be found in Table \ref{Table-AB} where pairs $(n_1,n_2)$ are $(1,-1)$, $(2,0)$ and $(0,-2)$ for the main SB, HH($m$), and HH($m-1$), respectively.


\begin{table}[H]
\begin{tabular}{|c|c|c|c|}
\hline
Peak & A & B & $\delta$\\ \hline

SB & 
$N^{2}\left[ F_{m-1}^{2}\left\vert I_{m-1}^{(1)}\right\vert^{2}+F_{m}^{2}\left\vert I_{m}^{(-1)}\right\vert ^{2}\right]$ 
& 
$2N^{2}F_{m-1}F_{m}\left\vert I_{m-1}^{(1)}\right\vert
\left\vert I_{m}^{(-1)}\right\vert$ 
&  
$\phi_{m}-\phi _{m-1}+\arg \Tilde{I}_{m-1}^{(1)} -\arg \left[\Tilde{I}_{m}^{(-1)}\right]$\\ \hline

HH($m$) & 
$N^{2}\left[ F_{m-1}^{2}\left\vert I_{m-1}^{(2)}\,\right\vert
^{2}+F_{m}^{2}\left\vert I_{m}^{(0)}\right\vert ^{2}\right]$ 
& 
$2N^{2}F_{m-1}F_{m}\left\vert I_{m-1}^{(2)}\,\right\vert \left\vert
I_{m}^{(0)}\right\vert$ 
& 
$\phi_{m}-\phi _{m-1}+\arg \Tilde{I}_{m-1}^{(2)} -\arg \left[\Tilde{I}_{m}^{(0)}\right]$\\ \hline

HH($m-1$) & 
$N^{2}\left[ F_{m-1}^{2}\left\vert I_{m-1}^{(0)}\right\vert
^{2}+F_{m}^{2}\left\vert I_{m}^{(-2)}\right\vert ^{2}\right]$
&
$2N^{2}F_{m-1}F_{m}\left\vert I_{m-1}^{(0)}\right\vert
\left\vert I_{m}^{(-2)}\right\vert $
& 
$\phi_{m}-\phi _{m-1}+\arg \Tilde{I}_{m-1}^{(0)} -\arg \left[\Tilde{I}_{m}^{(-2)}\right]$\\ \hline

\end{tabular}
\caption{Parameters for the transition probability given in Eq. (\ref{T-fit}) for the main SB and HH considered in this work.}
\label{Table-AB}
\end{table}


Fig. \ref{beta Savg} allows us to clarify the independence of $\Tilde{I}_{m}^{(n)}$ from the relative phase $\phi$ given by Eq. (\ref{T-appendix-1}). The key to understand the former statement arises from the observation that in the cases when the quantity $\beta _{+,m}^{(n)}$ changes its sign (absorption paths), it produces discontinuities in the factor $\cos {\left( \frac{\Delta S}{2}-\frac{\pi }{4}\text{\textrm{sgn}}[\beta _{+,m}^{(1)}]\right) }$.
On the other hand, the changes of sign in $e^{i\bar{S}_m}$ cancel the previous ones,
resulting in a continuous function $\Tilde{I}_{m}^{(n)}$ as a function of $F_{0}$,
as observed in Fig. \ref{Imn tilde} with the same color code as in (a).

\section{Classical and non-classical regions}
\label{Regions}

Whereas the classical region of each contribution can be ascribed to real ionization times given by Eqs. (\ref{t1}) and (\ref{t1p}), in the classical forbidden region the ionization times are complex as given by Eq. (\ref{stcomplex-2}). In the latter, as the real part of the complex time is constant, then $\bar{S} = S$ and $\Delta S = 0$.
Thus, $\tilde{I} \simeq e^{i \bar{S}}\frac{\sqrt{2}}{2}$. 
Overall, there are three types of energy regions for the RABBIT protocol:
(i) The classical region where both $T_m$ and $T_{m-1}$ in Eq. (\ref{T-RABBIT}) has real ionization times, (ii) the non-classical region where both $T_m$ and $T_{m-1}$ has complex ionization times, and (iii) the mixed region where one of the two contributions has real ionization times and the other has only one complex ionization time. The three regions are depicted in the scheme of Fig. \ref{complex-plots-limits}.

In the non-classical region, the only important factor for the interference pattern is the exponential factor $e^{i \bar{S}}$ for each of the two contributions $T_m$ and $T_{m-1}$ in Eq. (\ref{T-RABBIT}). 
Whereas the imaginary part of the complex action reveals itself as an exponential
decay of the probability, the real part emerges as an interference pattern.
In this case, since $\bar{S}_{m}$ and $\bar{S}_{m-1}$ are propotional to $\omega t_s$ as shown in Appendix \ref{A-SD-HH}, $\mathrm{Re}(\bar{S}_{m})$ and $\mathrm{Re}(\bar{S}_{m-1})$ differ in 
$\mathrm{Re}(2\omega t_s) = \pi$ or $3\pi$, that is, a change of sign. 
Therefore, the phase delay in the non-classical region is $\delta = \pi$.
When the energy of the marginal sidebands is such that it is in the mixed region one contribution has oscillating values of $\Tilde{I}$ (classical region) whereas the other contribution is in its classically forbidden region with
constant $\mathrm{Re}(\Tilde{I})$, as shown in Fig. \ref{Imn tilde}d. 
There are certain regions in the $F_0$ domain for which both contributions have the same sign and thus the phase delay is $\delta = 0$ (white shade), whereas there are other ones where they have different signs and thus $\delta$  = $\pi$ (yellow shade).
For low values of the probe field $F_0$, the marginal sidebands will be in opposition of phase with respect to the main SB and in phase with HH($m$) and HH($m-1$).


To get the classical limits, the transcendental equation $\left\vert \beta _{+,m}^{(n)}\right\vert=F_{0}/\omega$, must be numerically solved. However we can find an approximation for the case of very weak probe pulses (small values of the fundamental peak field $F_{0}$). In the low field strength regime, we can neglect the ponderomotive energy $U_{p}$ from the second term of $\beta _{+,m}^{(n)}$ in Eq. (\ref{saddle-s}), and find an analytical expression for the lower limit of validity of the SCM, i.e.,
\begin{equation}
F_{0}^{\min }=\omega \left\vert \sqrt{2n\omega +v_{0,m}^{2}}%
-v_{0,m}\right\vert .  \label{cl-val}
\end{equation}
%

\begin{acknowledgments}
This work is supported by PICT 2020-01755, 2020-01434, and PICT-2017-2945 of ANPCyT (Argentina), PIP 2022-2024 11220210100468CO of Consejo Nacional de Investigaciones Científicas y Técnicas (CONICET) (Argentina) and Programa Nacional RAICES Federal: Edición 2022 of Ministerio de Ciencia, Tecnología e Innovación (MinCyT).

\end{acknowledgments}

\bibliography{bibliografia-revised}

\end{document}